\documentclass[a4paper,fleqn,review]{cas-sc}

\usepackage[authoryear,longnamesfirst]{natbib}
\graphicspath{{figs/}}
\usepackage{xcolor}
\usepackage{setspace}

\def\tsc#1{\csdef{#1}{\textsc{\lowercase{#1}}\xspace}}
\tsc{WGM}
\tsc{QE}
\begin{document}
	\let\WriteBookmarks\relax
	\def\floatpagepagefraction{1}
	\def\textpagefraction{.001}
	
	\shorttitle{Rapid Identification Based DDTD}    
	
	\shortauthors{Yang et al.} 
	
	\title [mode = title]{ Rapid Identification Based Data-Driven Topology Design Independent from High-Information Entropy Initial Dataset}  
	
	\tnotemark[1] 
	
	\tnotetext[1]{%
		\textcopyright~2026. This manuscript version is made available under the CC-BY-NC-ND 4.0 license.\\
		\url{https://creativecommons.org/licenses/by-nc-nd/4.0/}\\
		The version of record is available at: \url{https://doi.org/10.1016/j.engstruct.2026.123810}
	} 
	
	
	\author[1]{Jun Yang}%[<options>]
	
	\cormark[1]
	
	
	\ead{yang\_jun2023@fuji.waseda.jp}
	
	\ead[url]{}
	
	\credit{Conceptualization, Methodology, Software, Writing – original draft}
	
	\affiliation[1]{organization={Graduate School of Information, Production and Systems, Waseda University},%Department and Organization
		addressline={2-7 Hibikino, Wakamatsu}, 
		city={Kitakyushu},
		postcode={808-0135}, 
		state={Fukuoka},
		country={JAPAN}}
	
	\cortext[1]{Corresponding author}
	
	\author[1]{Ziliang Wang}%[]
	
	% Footnote of the second author
	
	\credit{Visualization, Conceptualization}

	\author[1]{Shintaro Yamasaki}%[]
	\credit{Writing – review \& editing, Supervision, 
		Methodology}
	
	
	\begin{abstract}
		Topology optimization (TO) has been widely adopted in engineering design; however, it is prone to being trapped in local optima, particularly when addressing strongly nonlinear problems.
		Sensitivity-free data-driven topology design (DDTD) offers a promising alternative for such problems.
		Nevertheless, existing DDTD-based methods exhibit a strong dependence on prior information or sensitivity-based TO methods for initialization, which limits their generality and independence in engineering applications.
		Therefore, we establish an efficient DDTD-based framework capable of being driven from low information-entropy initial datasets while improving overall computational efficiency.
		To reduce the dependence of DDTD on high information-entropy initial datasets, a mesh-independent mutation module is introduced as a supplementary source of geometric features, enabling stable exploration under low information-entropy initialization.
		To address the computational bottleneck in DDTD, where all candidate structures require numerical evaluations, an embedding-based rapid identification algorithm is developed to efficiently select potential high-performance structures, significantly reducing the number of expensive high-fidelity simulations.
		The proposed framework generates material distributions on body-fitted meshes, ensuring consistency between numerical simulations and subsequent physical manufacturing. 
		To ensure reliable mesh generation, a signed distance field-based minimum length constraint is incorporated to improve mesh-generation stability and enhance structural manufacturability.
		Numerical experiments on strongly nonlinear stress-related problems, together with comparisons against sensitivity-based TO methods, validate the effectiveness of the proposed method. 
		In microfluidic reactor and shell design problems with non-differentiable constraints, the proposed method effectively addresses scenarios that remain challenging for sensitivity-based TO and conventional DDTD-based methods.

	\end{abstract}
	
	\begin{graphicalabstract}
		\includegraphics[width=0.86\textwidth]{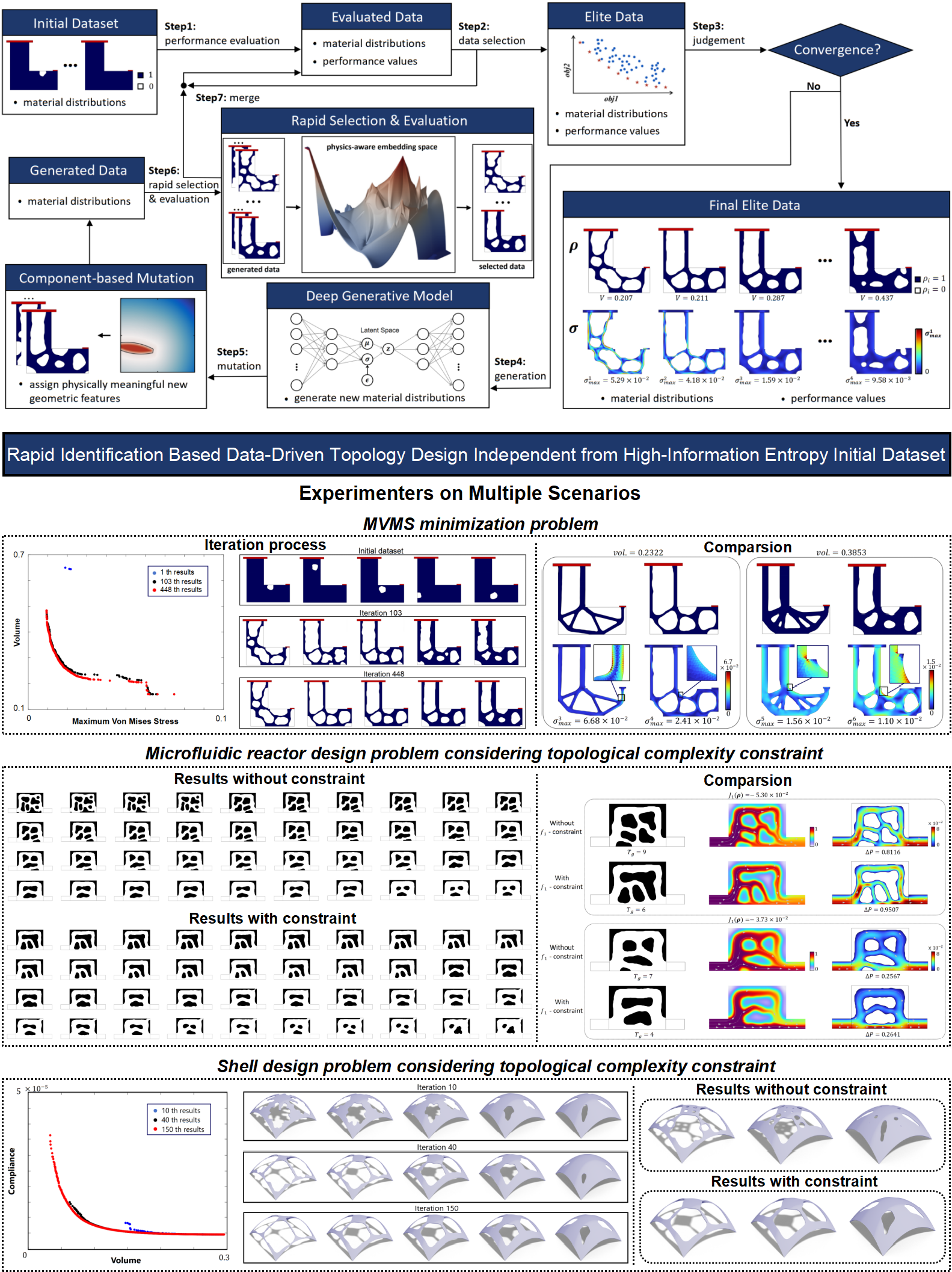}
	\end{graphicalabstract}

	\begin{highlights}
		\item A data-driven topology design framework driven by low-information entropy datasets
		\item Mesh-independent mutation module enhances robustness of global exploration
		\item Rapid identification algorithm reduces the number of costly high-fidelity evaluations
		\item SDF-based minimum length constraint improves meshing stability and manufacturability
		\item Effective for strongly nonlinear and non-differentiable optimization problems
	\end{highlights}

	% Keywords
	
	\maketitle
	
	% Main text
	\section{Introduction}
	\label{sec1}
	
	Topology optimization (TO) is a widely recognized computational methodology for generating high-performance structural designs by optimizing material distribution within a design domain. 
	Since its introduction by \cite{ref11}, TO has been extensively applied in numerous fields such as mechanics, aerospace engineering, and civil engineering (\cite{ref12,ref1,ref36,ref44}).

	Structural design based on TO has traditionally focused on stiffness maximization (\cite{ref12}).
	However, practical engineering applications often require explicit consideration of the maximum von Mises stress (MVMS) due to stress-induced fatigue and fracture failures (\cite{ref15}).
	Consequently, stress-related TO has attracted significant attention over the years (\cite{ref17,ref18,ref20}).
	Nevertheless, stress optimization remains challenging because of singularity, localization, and strong nonlinearity (\cite{ref20}).
	To improve numerical stability, conventional sensitivity-based TO methods usually employ stress-relaxation and approximation techniques (\cite{ref21,ref22,ref23}), which effectively transform the original stress objective into a modified differentiable formulation.

	With the development of sensitivity-free methods (\cite{ref25}), deep generative models have gradually been introduced into TO frameworks and demonstrated strong capability in handling large-scale design variables (\cite{ref28}).
	Based on this concept, \cite{ref29} proposed the sensitivity-free data-driven topology design (DDTD), where elite structures are iteratively selected, learned by deep generative models, and regenerated to progressively improve structural performance.
	The effectiveness of DDTD has been demonstrated in structural mechanics (\cite{ref31,ref55,ref54,ref57,ref61}), electromagnetics (\cite{ref58,ref52}), and fluid dynamics (\cite{ref14,ref30}), as well as practical engineering applications such as electromagnetic interference filters (\cite{ref52}), heat exchangers (\cite{ref30}), and thermal energy storage tubes (\cite{ref47}).

	A key limitation of these conventional DDTD-based methods lies in their intrinsic dependence on the quality of the initial dataset.
	Typically, reliable and high-performance results can only be achieved when a high-information entropy initial dataset is carefully prepared in advance.
	High-information entropy initial datasets generally require high-cost and time-consuming construction based on prior knowledge or existing TO methods, and therefore contain many high-performance structures. 
	In contrast, low-information entropy initial datasets emphasize low-cost and rapid construction, but their overall structural performance is typically low.
	By leveraging deep generative models through unsupervised learning, DDTD generates diverse geometric features while preserving key characteristics of the training data in the generation process.
	Nevertheless, the over-reliance on deep generative models exposes conventional DDTD-based methods to the same dilemma as deep generative models, i.e., the inability to learn meaningful features from low-information entropy initial datasets.
	This forces conventional DDTD-based methods to construct initial datasets with high-information entropy, in order to guarantee the performance of the final solutions.
	The high-information entropy initial dataset can be generated, e.g., by constructing a parametric model based on certain a priori information (\cite{ref52}), or by solving a pseudo-optimization problem that approximates the original optimization problem but is more easily solvable by using the sensitivity-based TO methods (\cite{ref30}).
	This indicates that conventional DDTD-based methods still remain dependent on prior information or sensitivity-based TO methods.

	\begin{figure}%[!t]
		\begin {center}
		\includegraphics[width=0.9 \textwidth]{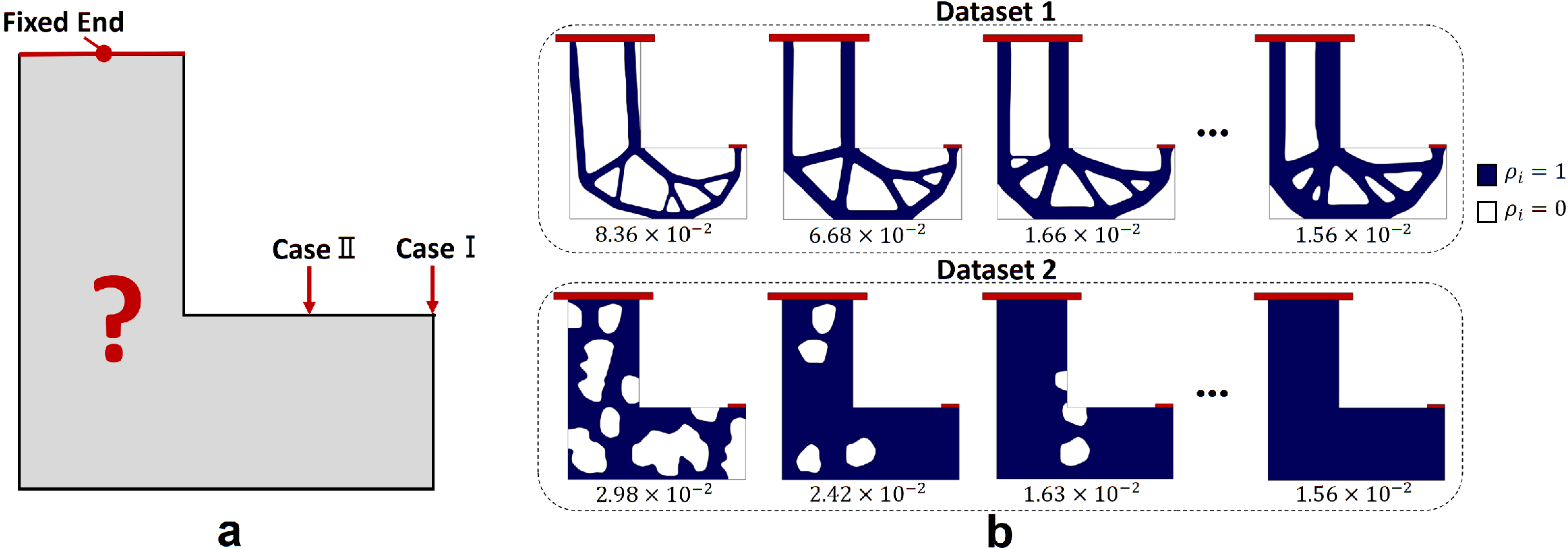}
		\caption{The impact of initial data with different information entropy on conventional DDTD-based methods. The value shown below each structure represents its maximum von Mises stress. }
		\label{highInfo}
		\end {center}
	\end{figure}

	Conventional DDTD-based methods are generally effective when initialized with high-information entropy datasets; however, constructing such datasets is computationally expensive and problem-dependent.
	As illustrated in Fig.~\ref{highInfo}(a), under load case~\uppercase\expandafter{\romannumeral 1}, dataset~1 containing high-performance structures enables effective DDTD optimization, whereas initialization from the low-performance dataset~2 leads to poor optimization performance.
	Moreover, datasets constructed for one optimization setting often lose effectiveness under modified boundary conditions (e.g., case~\uppercase\expandafter{\romannumeral 2}), indicating poor transferability.
	Although correlation-based mutation strategies have been introduced to alleviate this issue \cite{ref61}, their performance strongly depends on mesh-related hyper-parameters, limiting scalability for large-scale TO problems.
	Consequently, existing DDTD-based methods remain ineffective or impractical in low-information entropy scenarios.

	These limitations become more pronounced in practical TO problems involving discrete or non-differentiable characteristics, such as genus constraints.
	In 2D settings, the genus of a structure is equivalent to the number of enclosed cavities in this structure.
	Since genus is an inherently discrete topological quantity, conventional sensitivity-based TO methods can only impose indirect or relaxed constraints (\cite{ref_sup2,ref_sup3,ref_sup4,ref_sup5}), making exact topological control difficult.
	Furthermore, such problems usually lack reliable prior information, preventing sensitivity-based methods from generating high-quality initial solutions for DDTD initialization.
	As a result, conventional DDTD-based methods face fundamental limitations in these scenarios.

	More importantly, if the approach based on a mutation module is employed to start from a low-information entropy initial dataset, a huge number of trials is usually required for the mutation.
	This means that the computational cost will drastically increase when using this approach because the dominant portion of the computational cost in DDTD is the finite element analysis (FEA) on new structures generated by a deep generative model or the mutation module.
	As a result, existing DDTD-based methods face a computational bottleneck in the data-evaluation process, leading to increased computational demands in practical applications.

	These observations collectively indicate that the fundamental difficulty of applying conventional DDTD-based methods to scenarios lacking prior information lies in their intrinsic dependence on high-information entropy initial datasets. 
	In addition, existing DDTD-based methods suffer from substantial computational costs, a challenge that becomes increasingly pronounced in TO problems involving high-fidelity numerical simulation.
	In this paper, we therefore propose a rapid identification based DDTD-based method that can be driven by low-information entropy initial datasets, and apply it to strongly nonlinear engineering design problems, e.g., maximum stress minimization, and microfluidic reactor design problem with genus constraint.

	As a more reliable mutation module, we propose a parameter-controllable component-based mutation module that generates geometric features directly in the design domain.
	By utilizing a parameterized polygon as the basic geometric component, new topological variations are produced through a small number of controllable parameters. 
	All geometric operations are performed in the design domain, and the final assignment to the discrete design variables is achieved simply by checking whether the element centers lie inside the generated component.
	This mesh-independent mutation mechanism eliminates the need for repeated parameter retuning, significantly reduces computational overhead, and provides meaningful geometric features for deep generative models even in low-information entropy scenarios.

	To mitigate the computational cost bottleneck, we propose an embedding-based rapid identification algorithm that efficiently filters out potential low-performance structures before FEA, thereby reducing the overall computational burden. 
	High-performance structures generally exhibit similarities in geometry and topology, and structures that are close to each other in the high-dimensional solution space tend to show similar performance. 
	Based on this observation, we calculate the true structural performance of only a subset of the generated data using FEA and then project the high-dimensional solution space into a low-dimensional embedding space through the proposed physics embedding mapping. 
	The resulting physics-aware embedding space enables rapid identification of potential high-performance structures from the remaining performance-unknown dataset.
	In contrast to AI-based performance prediction methods, which require network architecture construction, hyper-parameter adjustment, and expensive retraining, the proposed embedding-based rapid identification algorithm incurs significantly lower computational overhead and provides strong adaptability.
	It achieves a practical balance between accuracy and computational efficiency, making it suitable for the proposed DDTD-based method.

	Additionally, we introduce a signed distance field (SDF)-based minimum length constraint to prevent the formation of excessively detailed outlines in the generated structures. 
	The primary purpose of this constraint is to guarantee stable and feasible generation of a body-fitted mesh, which is required to accurately evaluate the structural response on the true geometry of the design. 
	A secondary benefit of the constraint is that it improves the manufacturability of the resulting designs, ensuring that they satisfy the fundamental requirements of practical fabrication.

	The proposed DDTD-based method addresses the aforementioned limitations through the coordinated roles of its three components.
	The mesh-independent mutation module eliminates the dependence on high-information entropy initial datasets by generating diverse and meaningful geometric features, enabling robust exploration even under low-information entropy scenarios. 
	The embedding-based rapid identification algorithm reduces the overall computational burden by filtering out potential low-performance structures before numerical simulations, thereby enhancing efficiency.
	Furthermore, the SDF-based minimum length constraint ensures stable generation of body-fitted meshes, which is essential for accurate structural evaluation and manufacturability.

	The proposed method was applied to stress-related multi-objective optimization problems with strong nonlinear properties that may cause sensitivity-based TO methods to become trapped in local optima.
	Comparative experiments demonstrate the effectiveness of the proposed method. 
	Subsequently, we designed a microfluidic reactor design problem characterized by a topological complexity constraint. 
	Experimental results indicate that the proposed method can handle TO problems that are intractable for both sensitivity-based TO methods and conventional DDTD-based methods.
	Experiments in other optimization problems further demonstrate its effectiveness.

	In the following, the details of the proposed method are described in Section~\ref{sec2} and its effectiveness is confirmed using numerical examples in Section~\ref{sec3}. Finally, conclusions are provided in Section~\ref{sec4}.
	
	\section{Methods}
	\label{sec2}
	
	\subsection{Representation of material distributions and general formulation}
	\label{section_general_formulation}
	
\begin{figure}%[!t]
		\begin {center}
		\includegraphics[width=0.95 \textwidth]{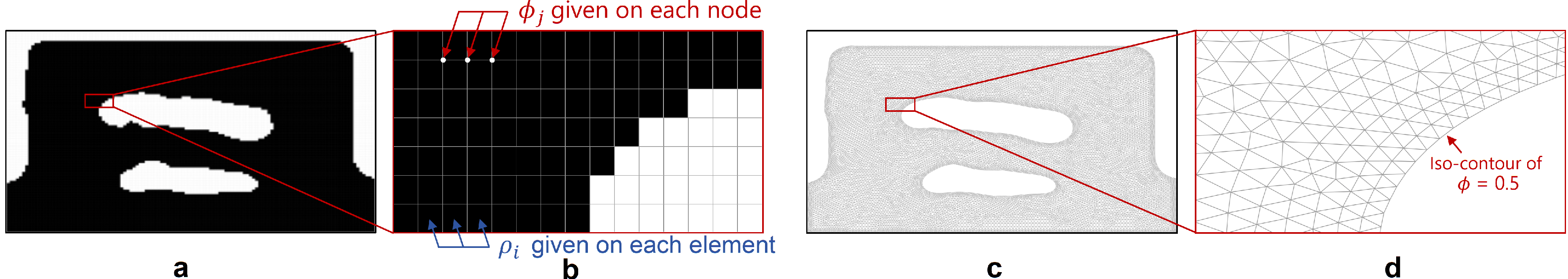}
		\caption{Design variables $\rho_i$ and corresponding smoothed material distribution $\phi$.}
		\label{fig_rho_phi}
		\end {center}
	\end{figure}

	In this section, we introduce the design variables to represent material distributions and explain a general formulation based on them.
	Let us consider a bounded and sufficiently regular domain $D \subset \mathbb{R}^{d}$ and ${d}\in{2,3}$, which defines the design domain for material distribution.
	Then, we discretize the design domain with a finite element mesh, and assign element-wise constant binary design variables $\rho_i \in \{0, 1\}$ $(i = 1, \dots, N_{\text{elem}})$ on each element.
	A convolution filter is applied to the design variables $\rho_i$ to obtain nodal variables $\phi_j$ as follows:
	\begin{equation}
		\phi_j = \frac{\sum^{N_{\text{elem}}}_{i=1} w_{i, j} \rho_i}{\sum^{N_{\text{elem}}}_{i=1} w_{i, j} }, \;\; \text{for} \; j=1, \dots, N_{\text{node}},
	\end{equation}
	and
	\begin{equation}
		\label{eq2}
		w_{i, j} = \max\left(0,\frac{r^{1}-r_{i,j}}{r^{1}}\right).
	\end{equation}
	where $r_{i, j}$ is the distance between the $j$-th node and the centroid of the $i$-th element, and $r^{1}$ is the filter radius.
	$N_{\text{elem}}$ and $N_{\text{node}}$ are the numbers of elements and nodes in the design domain, respectively.
	Using $\phi_j$ together with the FEM shape functions, we construct a field function $\phi(\mathbf{x})$ representing the presence of material at an arbitrary point $\mathbf{x}$ in $D$.
	Material is present at $\mathbf{x}$ if $\phi(\mathbf{x}) \ge 0.5$, and absent otherwise.
	An example is shown in Figure~\ref{fig_rho_phi}.
	Here, in Figure~\ref{fig_rho_phi}(a)-(b), the elements in black indicate $\rho_i = 1$ and the elements in white indicate $\rho_i = 0$.
	The convolution filter is applied to these $\rho_i$ to obtain $\phi_j$.
	The iso-contour of $\phi = 0.5$ is then extracted to obtain the material distribution represented by a body-fitted mesh as shown in Figure~\ref{fig_rho_phi}(c)-(d).
	This approach ensures the smoothness of the resulting material distribution by leveraging the convolution filter while introducing the element-wise constant and binary design variables.

	On the basis of the material distribution representation method explained above, we formulate a general multi-objective TO problem over $D$ as follows:
	\begin{equation}
		\label{eq0}
		\begin{array}{ll}
			\underset{\boldsymbol{\rho}}{\text{Minimize}} & \; \left[J_{1}(\boldsymbol{\rho}), \cdots, J_{N_{\text{obj}}}(\boldsymbol{\rho}) \right] \\ \\
			\text{Subject to} & \; f_k(\boldsymbol{\rho}) = 0, \;\; \text{for} \; k=1, \dots, N_{\text{ec}}, \\ \\
			& \; g_l(\boldsymbol{\rho}) \leq 0, \;\; \text{for} \; l=1, \dots, N_{\text{iec}}, \\ \\
			& \;  \rho_i \in\{ 0, 1 \}, \;\; \text{for} \; i=1, \dots, N_{\text{elem}} .
		\end{array}
	\end{equation}
	where $\boldsymbol{\rho}$ denotes the design variable vector.
	$J_*(\boldsymbol{\rho})$, $f_*(\boldsymbol{\rho})$ and $g_*(\boldsymbol{\rho})$ denote the $*$-th optimization objective, equality constraint and inequality constraint functions, respectively.
	Here, $N_{\text{obj}}$, $N_{\text{ec}}$ and $N_{\text{iec}}$ indicate the numbers of objectives, equality and inequality constraints, respectively.
	
	To calculate the structural responses for given material distributions, a numerical simulation framework based on FEA is adopted.
	That is, we generate a body-fitted mesh along with the iso-contour of $\phi = 0.5$ for a given material distribution and conduct FEA to evaluate the structural performance, i.e., $J_*(\boldsymbol{\rho})$, $f_*(\boldsymbol{\rho})$ and $g_*(\boldsymbol{\rho})$.
	In this way, the influence of spatial variations in material distribution on mechanical behavior can be quantitatively captured. 
	Consequently, FEA provides a solid foundation for correlating the design variables with physical responses, ensuring the data-driven process reflects the true structural performance.

	\subsection{Data processing workflow}
	\label{Data process flow}
	
	%%%\textcolor{blue}{
		%%%As a sensitivity-free methodology, DDTD-based methods employ deep generative models to guide the search for high-performance structural designs. 
		%%%The iteration process is composed of three principal stages: (i) evaluating and selecting high-performance structures as elite data, (ii) utilizing the selected elite data to train a generative model capable of extracting valuable features, (iii) generating candidate data with new diverse features and incorporating them into all the data. 
		%%%The overall structural performance of the elite data is continuously increased until it satisfies the convergence criterion.
		%%%By virtue of the property of tending to search globally from the solution space, DDTD is able to circumvent local optimal solutions and consequently offers high potential in solving strongly nonlinear optimization problems.
		%%%}
	
	\begin{figure}%[!t]
		\begin {center}
		\includegraphics[width=1 \textwidth]{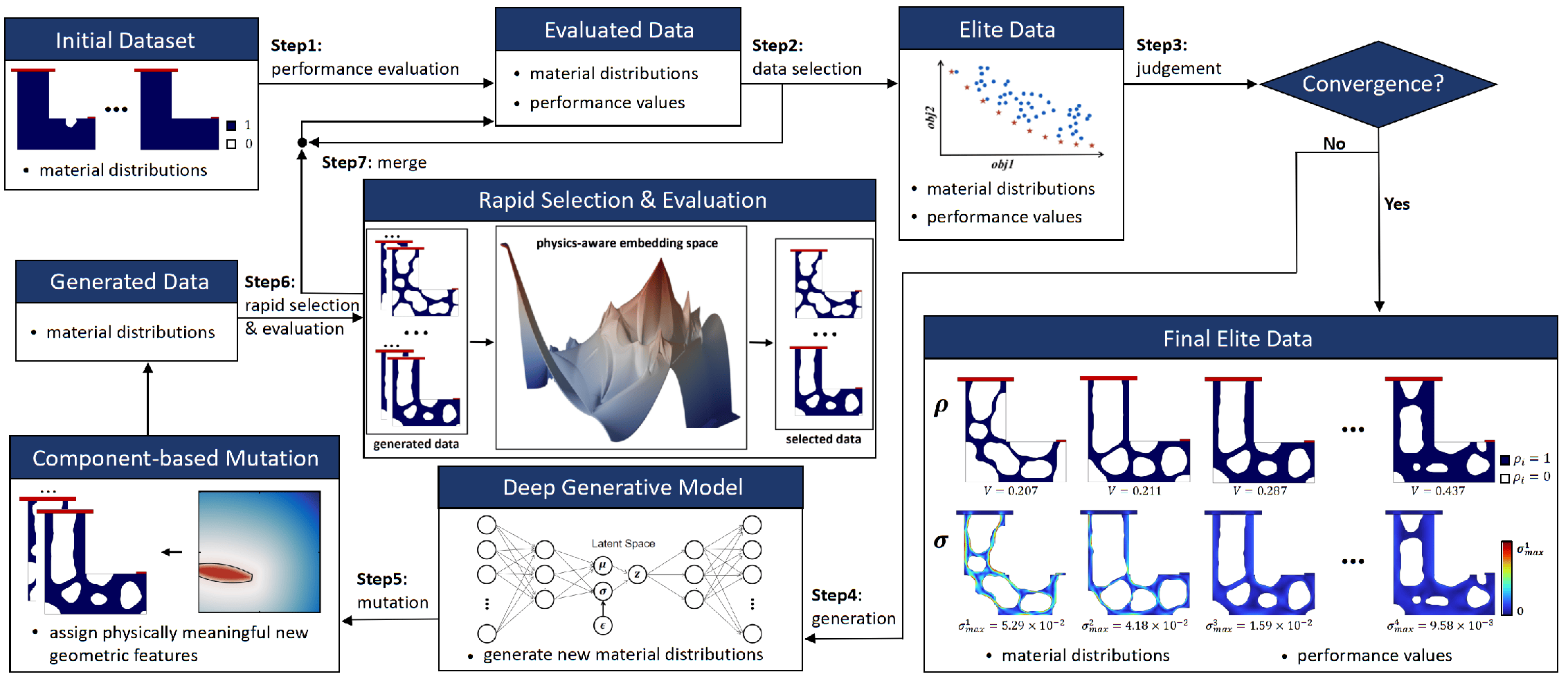}
		\caption{Data processing workflow of the proposed DDTD-based method.}
		\label{flowchart}
		\end {center}
	\end{figure}
	
	To solve optimization problems based on the general formulation explained in Section~\ref{section_general_formulation}, we construct the data processing workflow as shown in Figure~\ref{flowchart}.
	Initially, an initial dataset is prepared, which can be accomplished in various ways, e.g., introducing artificial parametric models (\cite{ref52}) or solving a pseudo-problem (\cite{ref30}).
	The preparation process determines the information entropy of the initial dataset with respect to the given optimization problem.
	A high-information entropy dataset generally can accelerate convergence; however, it often comes at the cost of increased computational overhead and reduced generalizability.
	
	After the initial setup, steps~1 through 7 are executed iteratively until a predefined convergence criterion is met.
	In Step 1, the structures in the initial dataset are evaluated, typically using FEA to assess the structural performance of each material distribution. 
	Step 2 applies a data selection algorithm to select high-performance structures, which serve as elite data.
	Step 3 evaluates the convergence using a metric based on the area outside of the elite data, which is a convergence indicator evaluating the overall performance of the elite data (\cite{ref29}). 
	If the convergence criterion is satisfied, the iteration terminates; otherwise, the process proceeds to Steps 4 and 5, which involve data generation and mutation.
	In Step 4, the elite data from Step 3 are employed to train a deep generative model, where the extracted features are compressed into a latent space. 
	Sampling in the latent space generates diverse material distributions while inheriting the original features of the training dataset. 
	Subsequently, the parameter-controlled component-based mutation module assigns new geometric features to the elite data, thereby releasing the over-reliance on deep generative models.
	This module provides informative features to the deep generative models in low-information entropy scenarios and enhances DDTD's capability to be driven from a low-information entropy initial dataset.
	In Step 6, potential high-performance structures are identified from candidate data (composed of the data generated by deep generative model and the mutated data) using the rapid selection module and their true structural performance is evaluated using FEA.
	By using the rapid selection module, we can significantly reduce computational cost in the scenario of the low-information entropy initial dataset.
	Step 7 merges the potential high-performance structures with the current elite data. 
	The merged data serve as the evaluated data of the next iteration.
	The details of each step are explained as follows.

	\textbf{Initial data generation: } 
	In conventional DDTD-based methods, a high-information entropy dataset is crucial to ensure the quality of the final solutions.
	However, the construction of high-information entropy datasets implies more expensive computation and demands sufficient a priori information.
	Variations in the optimization problem can cause the invalidation of the existing high-information entropy datasets.
	Additionally, conventional DDTD-based methods struggle to deal with scenarios where insufficient a priori information exists or where it is difficult to construct approximate yet easy to solve pseudo-problems, further limiting its applicability.
	To minimize the computational overhead, we construct the low-information entropy initial dataset using the mutation module described in Section~\ref{subsec_sdf}. 
	Specifically, one structure is first set as a fully solid material distribution in which all element densities are equal to $1$. 
	Then, a set of random geometric mutation operations, implemented using the mutation module, are applied to remove local material regions from the solid structure. 
	In this manner, a low-information entropy initial dataset can be efficiently constructed.
	Notably, the proposed method is not limited to this initialization strategy.

	\textbf{Performance evaluation and selection:}
	In this study, we evaluate the obtained material distribution data via FEA to calculate the value of the optimization objectives within the considered multi-objective optimization problem.
	Subsequently, we perform non-dominated sorting to select the rank-one material distributions as the elite data.
	In multi-objective optimization, the non-dominated sorting ranks solutions based on Pareto dominance, ensuring that the selected elite data represent optimal trade-offs among conflicting objectives. 
	Given a set of evaluated material distributions $ \boldsymbol{\hat{\rho}} =\left\{   \boldsymbol{\rho}^{(1)}, \boldsymbol{\rho}^{(2)}, \ldots, \boldsymbol{\rho}^{(N_{\mathrm{sol}} )}  \right\}$ ($N_{\mathrm{sol}}$ is the number of the evaluated material distributions) with optimization objective values $\left[J_{1}, J_{2}, \ldots, J_{N_{\mathrm{obj}}}\right]$,
	a solution $\boldsymbol{\rho}^{(i)}$ is considered to dominate another solution $\boldsymbol{\rho}^{(j)}$ if it satisfies: 
	\begin{equation}
		\label{pareto}
		\begin{aligned}
			&\forall k \in \left\{1, \ldots, N_{\mathrm{obj}}\right\}, \quad
			J_{k}\left( \boldsymbol{\rho}^{(i)} \right)
			\leq
			J_{k}\left( \boldsymbol{\rho}^{(j)} \right), \\
			&  \text{and}  \quad \exists k \in \left\{1, \ldots, N_{\mathrm{obj}}\right\}
			\text{ such that }
			J_{k}\left( \boldsymbol{\rho}^{(i)} \right)
			<
			J_{k}\left( \boldsymbol{\rho}^{(j)} \right).
		\end{aligned}
	\end{equation}
	
	Based on this criterion, the algorithm identifies the first non-dominated front, which consists of the rank-one material distributions.
	Here, we limit the maximum number of the elite data to $N_{\mathrm{eli}} $.
	When the number of the rank-one material distributions exceeds $N_{\mathrm{eli}}$, the elite data are chosen based on their crowding distance within the objective function space, as described by~\citep{ref33}.
	Otherwise, all the rank-one material distributions are considered as elite data.

	If the given convergence criterion is satisfied, we terminate the iteration process and obtain the current elite data as the final results. 
	Specifically, we define the convergence criterion as follows: if the change in the area outside of the elite data (a convergence indicator evaluating the overall performance of the elite data \citep{ref29}) is below a predetermined threshold for $N_{\mathrm{conv}}$ consecutive iterations, or if the iteration count reaches the maximum iteration number, the optimization process is terminated.
	The details of calculating the area outside of the elite data for each iteration are introduced in \cite{ref29}.

	\textbf{Generation of diverse new data:}
	The core of DDTD is to generate diverse new data according to the existing elite data, thereby providing candidate data with superior structural performance.
	In conventional DDTD-based methods, the generation process of diverse new data relies solely on the deep generative models, e.g., variational autoencoder (VAE) in this study. 
	Specifically, a fixed number of training data composed of elite data (with duplication when the number of elite data is insufficient) are used as input of the VAE to learn their features.
	After training, the VAE produces generated data that inherits the original features of the input data while acquiring diversity.
	However, deep generative models struggle to learn meaningful features from a low-information entropy initial dataset.
	This limitation is the primary reason why the conventional DDTD-based methods require the construction of a high-information entropy dataset.
	Therefore, mutation operations are applied to the elite data to enrich their geometric diversity, thereby mitigating the over-reliance on deep generative models.
	
	In DDTD driven by a low-information entropy initial dataset, the optimization process is divided into two phases: early stage and later stage.
	In the early stage, the geometric and topological features contained in the elite data are typically very limited, and the improvement of elite data primarily depends on the diversity introduced by mutation operations.
	In this mutation-dominated stage, multiple mutation operations are performed on the same elite data to further increase the feature diversity, and the amount of mutation-derived data is set to a large value (e.g., 27 times the number of structures generated by the VAE).
	This can quickly raise the performance of elite data from low-information entropy to high-information entropy data with a rough contour close to the optimal solutions.
	In the later stage, the features contained in the elite data exhibit certain regularities, enabling the deep generative model to effectively learn the intrinsic characteristics of the input data.
	Therefore, the VAE becomes the primary driving force for the solution exploration in the later stage.
	In this VAE-dominated stage, the amount of mutation-derived data is set to a small value (e.g., 1/8 of the number of structures generated by the VAE).
	The transition conditions between the two stages are defined as follows: if the change in the area outside of the elite data is below a predetermined threshold for $N_{\mathrm{trans}}$ consecutive iterations.

	\textbf{Rapid selection module:}
	The process of performing FEA calculations on all generated data typically incurs a substantial computational cost, which constitutes the primary computational bottleneck of DDTD-based methods.
	A straightforward way to mitigate this issue is to employ AI models to predict structural performance. 
	Nevertheless, the network construction, parameter tuning, and iterative training of AI models also impose considerable computational overhead. 
	More importantly, the predicted structural performance values obtained from AI models cannot be considered equivalent to the true structural performance values computed via FEA, as notable discrepancies often exist. 
	This deviation underscores that FEA remains an indispensable tool for obtaining true structural performance evaluations.
	Furthermore, as the elite data in the DDTD process are continuously updated, the geometry and topology of the generated data change significantly across iterations.
	This indicates that even though AI models are trained in one iteration of DDTD, it is still essential to retrain them at each iteration to adapt to structural changes, which further increases computational load.
	To address this challenge, and as introduced in Sections~\ref{embeddingMap} and \ref{rapid ident}, we propose an embedding-based rapid identification (RI) algorithm designed to efficiently select potential high-performance structures. 
	Compared to AI-based alternatives, this algorithm offers low computational cost and satisfactory identification efficiency, making it particularly well-suited for integration into DDTD frameworks.
	%%%To ensure the resulting structures conforms to fundamental requirements of practical manufacturing, we propose a signed distance field-based minimum length constraint that enables efficient evaluation of the minimum feature size within the target structure as introduced in Sect.~\ref{minLength}.
	%%%Data that do not satisfy the manufacturing constraints (including minimum length and connectivity constraint in this study) are not selected into potential high-performance structures, which further avoids frequent FEA evaluations.

	Through the above steps, we establish a DDTD-based framework that can be driven by a low-information entropy initial dataset, while the incorporation of the rapid selection module avoids frequent numerical evaluations, significantly enhancing the generality and efficiency.

	\subsection{Generation and mutation module}
	\label{subsec1}

	DDTD utilizes deep generative models to generate diverse data that differ from the training data.
	Nevertheless, DDTD's over-reliance on deep generative models makes it highly sensitive to the quality of the initial dataset.
	To address this limitation, we integrate the parameter-controllable component-based mutation module into the original DDTD methodology to assign geometric features to the input structures.
	In this study, the geometric features is defined as a parameter-controlled component with certain randomness, whose design is controlled by a subset of parameters.
	It is intended to meet specific mechanical or functional requirements by changing the shape and topological relationships of the original structure. 
	These geometric features introduce structural diversity.
	
	\begin{figure}%[h]
		\begin {center}
		\includegraphics[width=1 \textwidth]{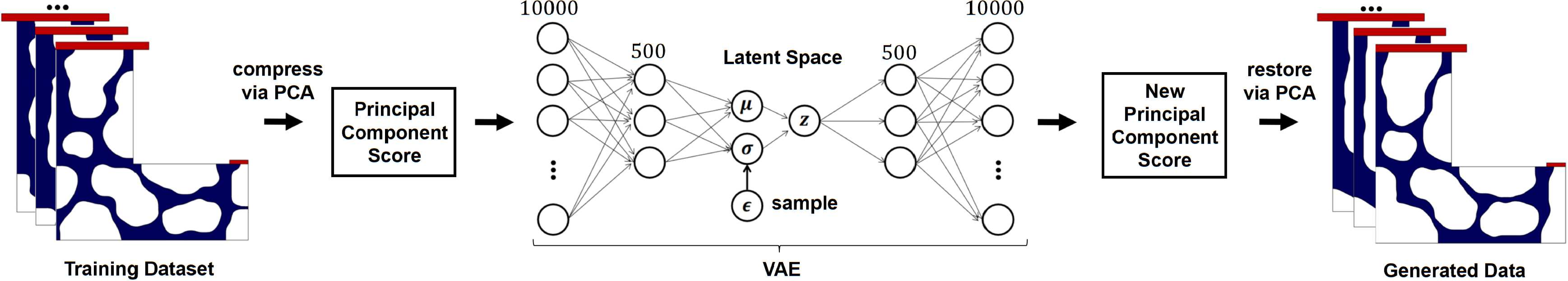}
		\caption{PCA-based VAE generation pipeline. }
		\label{fig_VAE}
		\end {center}
	\end{figure}
	
	\subsubsection{Generation module based on deep generative model}
	
	In the DDTD process, the training dataset is fed into the VAE, resulting in newly generated data. 
	The number of material distributions in the training dataset is set to 
	$N_{\mathrm{eli}}$. 
	The elite data are directly used as the training dataset when the number of elite data equals $N_{\mathrm{eli}}$. 
	Otherwise, data augmentation is performed by replicating the elite data.
	
	The detailed generation process based on VAE follows the principal component analysis (PCA)-based DDTD framework proposed in~\cite{ref55}. 
	The PCA-based VAE generation pipeline is shown in Figure~\ref{fig_VAE}.
	The encoder of VAE consists of two hidden layers with 10000 and 500 neurons, while the decoder adopts a symmetric architecture with 500 and 10000 neurons. 
	ReLU activation functions are used in the hidden layers, and the VAE is trained using a loss function consisting of a reconstruction loss and a Kullback–Leibler divergence term. 
	Additional implementation details are identical to those reported in~\cite{ref55}.
	Here, we first generate material distributions including grayscale elements from the element-wise constant and binary design variables, by using an element-to-element convolution filter.
	Subsequently, the material distributions are converted into a principal component score matrix using PCA. The VAE trained on this matrix then generates a new principal component score matrix, and new material distributions are restored via the inverse PCA mapping.
	Finally, by applying a threshold of 0.5, we obtain the set of the design variables corresponding to the newly generated material distributions.

	%% Use \subsection commands to start a subsection.
	\subsubsection{Parameter-controlled component-based mutation module}
	\label{subsec_sdf}

\begin{figure}%[h]
		\begin {center}
		\includegraphics[width=1 \textwidth]{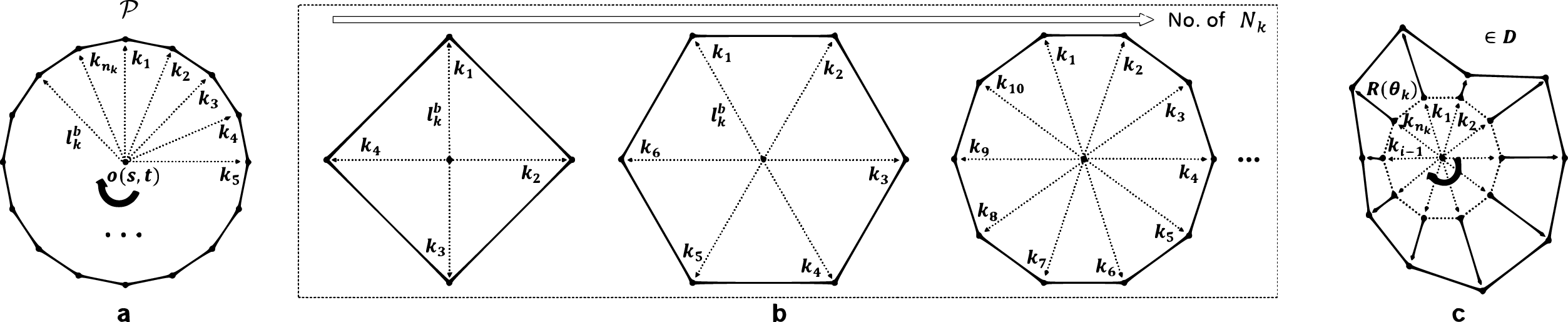}
		\caption{Partial shape of parameter-controllable polygon by random parameters. }
		\label{fig_polygon}
		\end {center}
	\end{figure}

	The purpose of the mutation module is to add geometric features with certain randomness to the input material distribution.
	Let us consider a two-dimensional example, i.e., $D \subset \mathbb{R}^{d=2}$.
	The material distribution and mutation operations are all located in $D$.
	The geometric features generated by the mutation module are inserted into the material distribution via projection onto the design domain.
	In $D$, the mutation module generates geometric features with a certain randomness by performing mutation operations. 
	In this study, we define this geometric feature as a parameter-controllable polygon $\mathcal{P}$, represented as follows:
	
	\begin{equation}
		\label{eq_comp}
		\mathcal{P} = 
		\bigl\{
		\mathbf{f}_k(s, t)
		= \mathbf{o}(s,t)
		+ R_k
		\begin{bmatrix}
			\cos(\theta_k)\\[2pt]
			\sin(\theta_k)
		\end{bmatrix},
		\quad k = 0,1,\dots,N_{\mathrm{k}} -1
		\bigr\}.
	\end{equation}
	where, as shown in Figure~\ref{fig_polygon}(a), $\mathcal{P}$ is a closed polygon defined by a center point and $N_{\mathrm{k}}$ radii extending in different directions. 
	%%%$\tau$ is random rotation angle.
	A center point $\mathbf{o}(s, t)$ is randomly selected in $D$, and $N_{\mathrm{k}}$ radii are emitted at equal angles around it. 
	By assigning a specific length to each radius, the positions of the boundary points are determined, ultimately forming a closed polygon that serves as a local mutation region over $D$.
	The length of each radius $R_k$ is randomly determined in the range of the pre-defined maximum length and minimum length.
	By changing the value of $R_k$ randomly, the shape of the parameter-controllable polygon can be altered to produce diverse geometric features as shown in  Figure~\ref{fig_polygon}(c).
	
	The parameter $N_{\mathrm{k}}$ determines the shape complexity of parameter-controllable polygon as shown in Figure~\ref{fig_polygon}(b) and the angle value $\theta_{k}$ for each ray is calculated as follows:
	
	\begin{equation}
		\label{eq_zeta}
		\theta_{k}=\frac{2 \pi k}{ N_{\mathrm{k}} }, \quad k=0,1, \ldots,  N_{\mathrm{k}}-1.
	\end{equation}

	Using the polygon defined in Eq.~\ref{eq_comp}, we mutate the material distribution.
	That is, we detect elements whose center point is in or on the polygon and change their densities to 0.
	By doing so, we add a lump of void to a given material distribution.
	On the other hand, we can add a lump of solid to a given material distribution by changing these densities to 1.
	The parameter-controllable polygon is adopted in the mutation module to represent geometric features, owing to its inherent advantages in both clear geometric representation and low computational cost.
	Furthermore, the mutation operations maintain a high geometric fidelity while without depending on a fixed structured mesh.
	Their robustness in performing geometric Boolean operations such as union, intersection, and subtraction can also support efficient shape blending and constructive modeling within the mutation process (\cite{ref43}).
	Through the process of assigning geometric features, the shape and topology of the original structures are altered, thereby transforming the performance of the original structures.

	\subsection{Rapid selection module}
	
	The purpose of the rapid selection module is to improve the computational efficiency of DDTD by reducing the number of structures requiring FEM evaluations.
	This module consists of two components: a physics-embedded mapping algorithm and a rapid identification (RI) algorithm built upon it.
	The core idea of the RI algorithm is to first construct a low-dimensional embedding space $\boldsymbol{\xi}$ for a performance-unknown dataset $\boldsymbol{\hat{\rho}} =\left\{   \boldsymbol{\rho}^{(1)}, \boldsymbol{\rho}^{(2)}, \ldots, \boldsymbol{\rho}^{( N_{\mathrm{sol}}  )}  \right\}$ in a high-dimensional space, where the number of structures in $\hat{\boldsymbol{\rho}}$ is denoted by $N_{\mathrm{sol}}$.
	The dataset $\hat{\boldsymbol{\rho}}$ is divided into two disjoint subsets.
	Specifically, a realistic performance dataset
	$\hat{\boldsymbol{\rho}}_{\mathrm{real}}$
	is constructed by randomly selecting
	$N_{\mathrm{real}}$ data from
	$\hat{\boldsymbol{\rho}}$.
	These randomly selected data are evaluated using FEA to obtain their true performance values.
	The remaining $N_{\mathrm{val}}$ data in $\hat{\boldsymbol{\rho}}$ constitute the validation dataset
	$\hat{\boldsymbol{\rho}}_{\mathrm{val}}$, where $N_{\mathrm{real}} + N_{\mathrm{val}} = N_{\mathrm{sol}}$.
	Based on the positions of structures
	$\hat{\boldsymbol{\rho}}_{\mathrm{real}}$
	in the embedding space $\boldsymbol{\xi}$ and their associated true performance values, an interpolation algorithm implemented using the built-in MATLAB function 'scatteredInterpolant' is employed to estimate the performance values at arbitrary locations within $\boldsymbol{\xi}$, thereby constructing a physics-aware embedding space.
	
	The 'scatteredInterpolant' function generates a continuous interpolation field from irregularly distributed sample nodes. 
	Specifically, the coordinates of $\hat{\boldsymbol{\rho}}_{\mathrm{real}}$ in the embedding space are treated as sample nodes, and their associated true performance values are assigned to these nodes. 
	Natural neighbor interpolation is then used to estimate the performance values at unsampled locations, whereas linear extrapolation is applied outside the convex hull of the sampled nodes.

	Notably, the adopted interpolation scheme does not necessarily represent the optimal choice for the proposed framework, and further investigation of alternative interpolation strategies may provide additional improvements in future studies.
	
	Since high-performance structures tend to exhibit similar geometric and topological features, structures that are close in the original high-dimensional space often share similar performance characteristics. 
	This property is largely preserved in the relative layout of data within $\boldsymbol{\xi}$. 
	Subsequently, according to the position of the data in the validation dataset
	$\hat{\boldsymbol{\rho}}_{\mathrm{val}}$
	within $\boldsymbol{\xi}$, we identify data whose performance is likely to satisfy the prescribed criterion.
	FEM evaluations are performed only on the potential high-performance structures, while the remaining data in 
	$\hat{\boldsymbol{\rho}}_{\mathrm{val}}$
	are neglected.

	\subsubsection{Physics-embedded mapping algorithm} \label{embeddingMap}
	
	The physics-embedded mapping algorithm plays a critical role in the proposed framework, as it enables the mapping of material distributions with massive design variables into a low-dimensional embedding space while preserving essential high-dimensional geometric and topological information. 
	
	Variations in the design variables directly lead to changes in the geometric and topological configuration of the material distribution, thereby influencing the physical response of that structure.
	Consequently, the discrepancies in geometry and topology between various material distributions can be reflected in differences between their corresponding design variables, as follows:
	
	\begin{equation}
		\label{eqst}
		\begin{aligned}
			d_{i j}=\left\|\boldsymbol{\rho}^{(i)}-\boldsymbol{\rho}^{(j)}\right\|_{2}=\sqrt{\sum_{k=1}^{N_{\text{elem}}}\left({\rho_{k}}^{(i)}-{\rho_{k}}^{(j)}\right)^{2}}, \quad \text { for } 1 \leqslant i<j \leqslant N_{\text{sol}},
		\end{aligned}
	\end{equation}
	where, $d_{i j}$ represents the geometric and topological discrepancy between material distributions $\boldsymbol{\rho}^{(i)}$ and $\boldsymbol{\rho}^{(j)}$. 
	It reflects the degree of similarity between both structures characterized by the design variables and is regarded as the distance between them in a high-dimensional solution space $\mathbb{R}^{d=N_{\text{elem}}}$, where the dimensionality $N_{\text{elem}}$ corresponds to the number of design variables.
	The $\boldsymbol{\rho}^{(i)}, \boldsymbol{\rho}^{(j)}\in \mathbb{R}^{N_{\text{elem}} \times 1}$ belong to a dataset $\hat{\boldsymbol{\rho}} \in \mathbb{R}^{N_{\text{elem}} \times N_{\text{sol}} }$, which consists of $N_{\text{sol}}$ material distributions.
	We construct the pairwise Euclidean distance matrix $\mathbf{P}$ for $\hat{\boldsymbol{\rho}}$ from Eq.~\ref{eqst} and apply classical multidimensional scaling (MDS) (\cite{ref59}) to $\mathbf{P}$ with the embedding dimension $N_{\mathrm{es}}$, where it is significantly lower than the original dimension $N_{\text{elem}}$.
	By doing so, we obtain $\boldsymbol{\hat{\xi}} \in \mathbb{R}^{N_{\text{sol}} \times N_{\mathrm{es}} }$ that
	represents the positions of the data in the low-dimensional embedding space corresponding to the data in the original high-dimensional space.
	The position of $\boldsymbol{\hat{\rho}}$ in $\boldsymbol{\xi}$ is characterized by $N_{\mathrm{es}}$ embedding dimensions.

	\subsubsection{Rapid identification algorithm} \label{rapid ident}

	The dataset $\hat{\boldsymbol{\rho}}$ contains $m$ material distributions with unknown performance.
	Through the process introduced in Section~\ref{embeddingMap}, the $\hat{\boldsymbol{\rho}}$ in the high-dimensional solution space $\mathbb{R}^{d=N_{\text{elem}}}$ is projected onto a low-dimensional embedding space $\mathbb{R}^{d=N_{\mathrm{es}} }$ ($N_{\mathrm{es}} <N_{\text{elem}}$).
	To endow the embedding space with the ability to predict the structural performance according to the relative positions in $\boldsymbol{\xi}$, it is necessary to calculate the true performance values of partial data in $\hat{\boldsymbol{\rho}}$, that is realistic performance data, to construct a physics-aware embedding space.
	The true performance values of realistic performance data
	$\hat{\boldsymbol{\rho}}_{\mathrm{real}}$
	are typically evaluated by numerical methods. 
	The remaining structures constitute the validation dataset
	$\hat{\boldsymbol{\rho}}_{\mathrm{val}}$.

	According to the positions of structures
	$\hat{\boldsymbol{\rho}}_{\mathrm{real}}$
	in the embedding space and their true performance values
	$(J_1, J_2, \cdots, J_{N_{\text{obj}}})$, an interpolation algorithm is employed to estimate the performance values at arbitrary locations within $\boldsymbol{\xi}$, resulting in a physics-aware embedding space.
	The estimated performance of
	$\hat{\boldsymbol{\rho}}_{\mathrm{val}}$
	is noted as $\mathcal{F}\in \mathbb{R}^{N_{\mathrm{val} }  \times N_{\mathrm{obj}}}$, and each row represents the objective values of $i$-th data under $N_{\mathrm{obj}}$ optimization objectives.
	Subsequently, a selection criterion for identifying potential high-performance structures within
	$\hat{\boldsymbol{\rho}}_{\mathrm{val}}$
	is established, whereby data that do not satisfy this criterion are removed from
	$\hat{\boldsymbol{\rho}}_{\mathrm{val}}$.
	In this study, the selection criterion is specified as picking out $N_{\mathrm{phps}} $ potential high-performance structures based on their corresponding non-dominated rank and crowding distance~\citep{ref33}.
	Structures in 
	$\hat{\boldsymbol{\rho}}_{\mathrm{val}}$
	that are not identified as potential high-performance structures are excluded from FEA evaluation, thereby significantly improving the overall computational efficiency.
	Therefore, within the dataset $\hat{\boldsymbol{\rho}}$, the proportion of data requiring FEA calculations is defined as the ratio of the combined number of potential high-performance structures and
	$\hat{\boldsymbol{\rho}}_{\mathrm{real}}$
	to the total number of $\hat{\boldsymbol{\rho}}$.

	\begin{figure}%[H]
		\begin {center}
		\includegraphics[width=1 \textwidth]{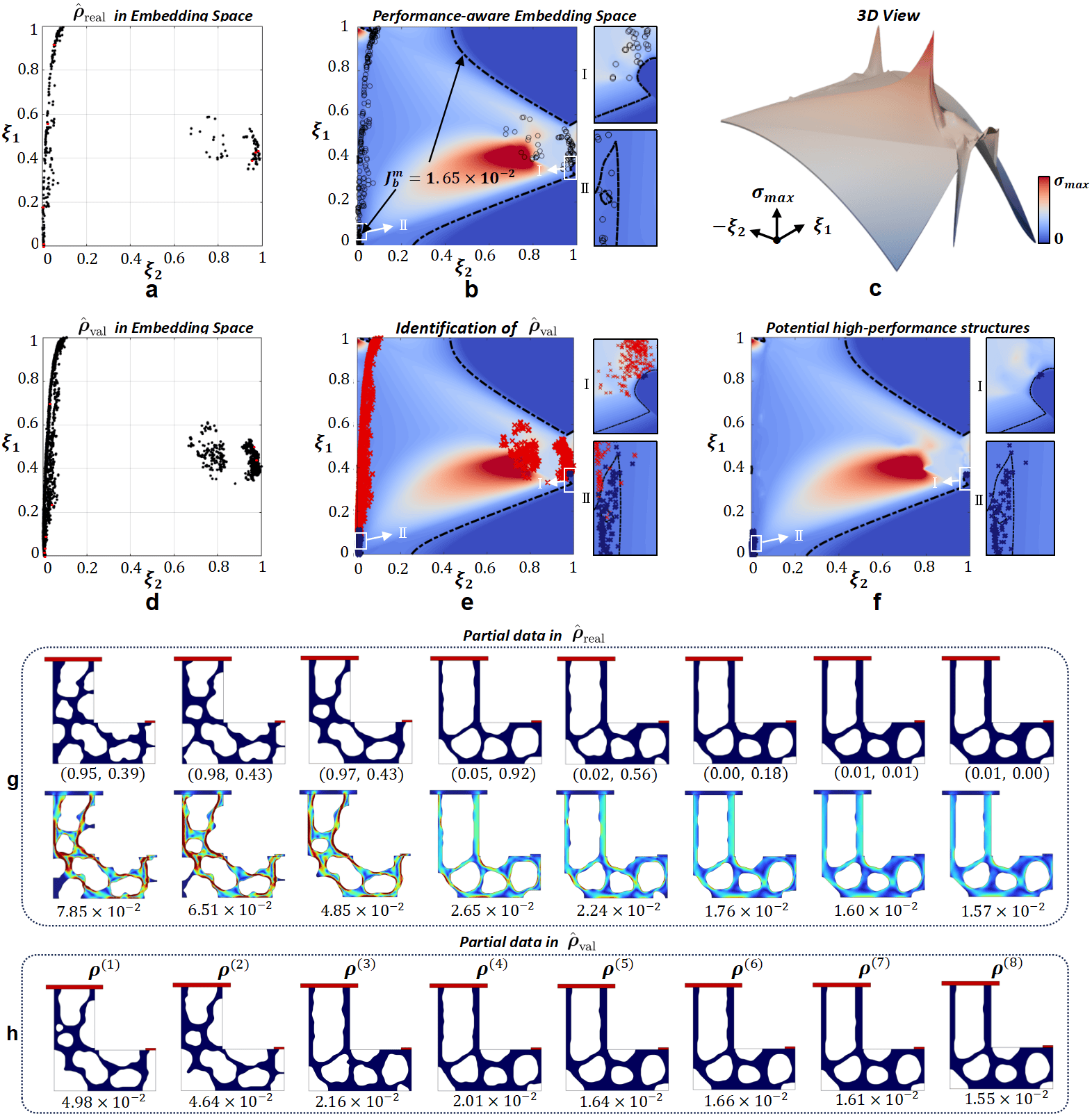}
		\caption{An example of rapid identification algorithm.}
		\label{fig_ex}
		\end {center}
	\end{figure}

	An example is shown in Figure~\ref{fig_ex}. 
	We construct a performance-unknown dataset $\hat{\boldsymbol{\rho}}$ consisting of $3848$ material distributions, each of which has $6400$ design variables.
	Using the physics-embedded mapping algorithm, the original high-dimensional design space $\boldsymbol{\rho} \in \mathbb{R}^{d=6400}$ is projected onto a two-dimensional embedding space $\boldsymbol{\xi} = (\xi_1, \xi_2)$. 
	The realistic performance subset
	$\hat{\boldsymbol{\rho}}_{\mathrm{real}}$
	including 300 material distributions are randomly selected from $\hat{\boldsymbol{\rho}}$, and their MVMSs are computed via FEA to serve as the true performance values. 
	(a) shows the normalized positions in $\boldsymbol{\xi}$ for all material distributions in dataset
	$\hat{\boldsymbol{\rho}}_{\mathrm{real}}$, and (g) shows partial material distributions (below are its coordinates in $\boldsymbol{\xi}$) and corresponding stress distributions (below are the MVMS values) in $\hat{\boldsymbol{\rho}}_{\mathrm{real}}$.
	Notably, the red dots in (a) and (d) highlight the positions of the structures shown in (g) and (h), respectively, within the embedding space.
	According to the positions and true performance values of structures in
	$\hat{\boldsymbol{\rho}}_{\mathrm{real}}$, we construct a physics-aware embedding space as shown in (b), and (c) is the 3D view of (b).
	We set $N_{\mathrm{phps} }=350$ and select $350$ potential high-performance structures from
	$\hat{\boldsymbol{\rho}}_{\mathrm{val}}$
	based on the interpolated structural performance values (the prediction values) via the normalized positions of
	$\hat{\boldsymbol{\rho}}_{\mathrm{val}}$
	in $\boldsymbol{\xi}$ as shown in (d), and the partial structures of $N_{\mathrm{phps}}$ data are shown in $\boldsymbol{\rho}^{(6)}$ to $\boldsymbol{\rho}^{(8)}$ (below are the true MVMS values via FEM calculations) in (h).
	A slightly higher value than the interpolated MVMS value of these 
	$N_{\mathrm{phps}}$ potential high-performance structures is then chosen as an observation threshold $J_b^{m}=1.65\times10^{-2}$, and its contour is plotted as illustrated in (b).
	Notably, the data that are not selected as the $N_{\mathrm{phps}}$ potential high-performance structures in
	$\hat{\boldsymbol{\rho}}_{\mathrm{val}}$
	are not subjected to FEA calculations in the proposed DDTD-based method, and the partial structures of potential low-performance structures are shown in $\boldsymbol{\rho}^{(1)}$ to $\boldsymbol{\rho}^{(5)}$ in (h).
	In this example, FEA calculations are applied to all the data in
	$\hat{\boldsymbol{\rho}}_{\mathrm{val}}$
	solely for validating the effectiveness of the proposed algorithm.
	Data in
	$\hat{\boldsymbol{\rho}}_{\mathrm{val}}$
	whose true structural performance values exceed this threshold $J_b^{m}$ are marked in red symbols in (e), whereas the remaining data are marked in blue symbols. 
	For clarity, (f) displays only the structures in
	$\hat{\boldsymbol{\rho}}_{\mathrm{val}}$
	whose true structural performance values are lower than this threshold $J_b^{m}$.
	As observed, structures with similar performance tend to cluster in nearby regions of the embedding space.

	Data that do not satisfy the selection criterion but are incorrectly identified as potential high-performance structures, e.g., $\boldsymbol{\rho}^{(6)}$ in (h), as well as those whose true performance satisfies the selection criterion but are not identified, e.g., $\boldsymbol{\rho}^{(5)}$ in (h), are regarded as selection-bias data.
	As shown in (e) and (f), these selection-bias data are mostly located near the observation threshold $J_b^{m}$. 
	The selection bias ratio on
	$\hat{\boldsymbol{\rho}}_{\mathrm{val}}$
	was $1.58\%$, corresponding to the identification accuracy of  $98.42\%$.
	The computational time for the physics-embedded mapping and rapid identification algorithms was 14.7 seconds and 0.1 seconds, respectively.
	This indicates that we only need to calculate the structural performance of $\hat{\boldsymbol{\rho}}_{\mathrm{real}}$ ( here, $ N_{\text{real}}=300$), and the $350$ potential high-performance structures identified from 
	$\hat{\boldsymbol{\rho}}_{\mathrm{val}}$
	using FEA, thereby circumventing the FEA calculation of the remaining $3198$ structures (accounting for $83.11\%$ of $\hat{\boldsymbol{\rho}}$) to reduce the computational overhead.

	It should be noted that the proposed rapid identification algorithm is not intended to provide precise performance prediction for individual structures, but rather to perform a preliminary screening among a large number of candidate structures. 
	The optimization process of DDTD relies on the iterative update and selection of the entire dataset rather than the prediction result of any single structure. 
	Although the Euclidean distance-based similarity measure may not fully capture the influence of local geometric variations on physical responses in certain cases, the overall effectiveness of the proposed framework is not significantly affected.
	For example, structures with a smooth fillet and those with a sharp corner may exhibit significantly different MVMS values despite having similar geometric representations.
	However, the impact of this limitation on the overall DDTD process is limited.
	On the one hand, due to the limitation of the similarity measure, some candidate structures may be incorrectly identified or missed during the rapid identification process, which leads to the occurrence of selection-bias data. 
	Nevertheless, potential high-performance structures selected via the RI algorithm will still undergo high-fidelity numerical evaluation, and their true performance will be accurately determined during the evaluation process. 
	Structures that fail to meet the performance criteria will therefore be eliminated during the selection stage.
	On the other hand, since the DDTD search process is driven by the iterative expansion and updating of the elite data rather than relying on a single identification step, the generative model and mutation module can still generate structures located near potential high-performance regions in subsequent iterations, allowing them to be identified and incorporated into the elite data during later stages.
	Therefore, the existence of a small amount of selection-bias data in individual iterations does not fundamentally alter the search direction or the final optimization results. 
	In essence, the primary role of the rapid identification algorithm is to reduce unnecessary numerical evaluations and improve computational efficiency, rather than to restrict the exploration capability or the ultimate optimization performance of the DDTD framework.

	\subsection{Minimum length and connectivity constraints}
	\label{minLength}

	As explained in Section~\ref{section_general_formulation}, a body-fitted mesh is used to evaluate the structural performance of the material distribution.
	In this case, body-fitted mesh generation may fail when the material distribution contains excessively complex boundaries.
	To avoid such a situation, we introduce a minimum length constraint based on the signed distance function (SDF), and material distributions violating this constraint are discarded before generating the body-fitted mesh.
	As a beneficial side effect, the manufacturability is ensured by this constraint.
	We also introduce a connectivity constraint to detect material distributions containing isolated solid domains in mechanical design problems, and eliminate these material distributions before generating the body-fitted mesh.

	Unlike conventional minimum length constraint methods that rely on explicit geometric measurements, the proposed minimum length constraint is based on a sequence of operations on the signed distance function of the material distribution.
	We start from the field function $\phi(\mathbf{x})$ whose iso-contour of $0.5$ represents the structural boundary (see Section~\ref{section_general_formulation}).
	We first make a signed distance function $\psi_{\text{org}}(\mathbf{x})$ as follows:
	\begin{equation}
		\psi_{\text{org} }(\mathbf{x}) = \mathcal{R}(2 \phi(\mathbf{x}) - 1),
	\end{equation}
	where $\mathcal{R}$ denotes the operation of a geometry-based re-initialization scheme proposed in ~\cite{ref60}. 
	The transformation $2 \phi(\mathbf{x})-1$ maps the material distribution to the range $[-1,1]$, enabling subsequent operations on the SDF.
	Next, we construct a reference field function $\psi_{\text{mod}}(\mathbf{x})$ that removes thin regions smaller than a specified minimum length $w_{\min}$ as follows:
	\begin{equation}
		\label{eq9}
		\psi_{\text{mod}}(\mathbf{x}) = \mathcal{R}\left(\psi_{\text{org}}(\mathbf{x})-\frac{w_{\min}}{2}\right)+\frac{w_{\min}}{2}.
	\end{equation}
	Here, the thin region smaller than $w_{\min}$ becomes less than $0$ in $\psi_{\text{org}}(\mathbf{x})-\frac{w_{\min}}{2}$ due to the mathematical properties of the SDF.
	Therefore, $\psi_{\text{org}}(\mathbf{x})$ and $\psi_{\text{mod}}(\mathbf{x})$ are identical except for the thin regions smaller than $w_{\min}$ in $\psi_{\text{org}}(\mathbf{x})$.
	In this study, $w_{\min}$ is typically set to twice the minimum finite element edge length.

	To quantify the difference between $\psi_{\text{org}}(\mathbf{x})$ and $\psi_{\text{mod}}(\mathbf{x})$, they are mapped to material distribution fields $H(\psi_{\text{org}}, h)$ and $H(\psi_{\text{mod}}, h)$, respectively, using a relaxed Heaviside function given as follows:
	\begin{equation}
		H(\psi, h)=\left\{\begin{array}{ll}
			0, & \psi \leq-h, \\
			\frac{1}{2}+\frac{\psi}{h}\left(\frac{15}{16}-\frac{5}{8}\left(\frac{\psi}{h}\right)^{2}+\frac{3}{16}\left(\frac{\psi}{h}\right)^{4}\right), & -h \leq \psi \leq h, \\
			1, & \psi \geq h ,
		\end{array}\right.
	\end{equation}
	where $h$ is the transition width.
	Then, we calculate the maximum difference between $H(\psi_{\text{org}}, h)$ and $H(\psi_{\text{mod}}, h)$ in $D$ as
	\begin{equation}
		\Delta H = \max_{D} \left\{ H\left(\psi_{\text{org}}, h\right)-H\left(\psi_{\text{mod}}, h\right) \right\}.
	\end{equation}
	$\Delta H$ is a measure of the discrepancy between $\psi_{\text{org}}(\mathbf{x})$ and $\psi_{\text{mod}}(\mathbf{x})$, and we judge that $\phi(\mathbf{x})$ violates the minimum length constraint if $\Delta H$ exceeds a given threshold $\tau$.

	For the connectivity constraint, isolated solid domains are detected by examining the adjacency relationships between elements in the design variable field $\boldsymbol{\rho}$.

	\section{Experiments and discussion}
	\label{sec3}

	This section demonstrates the application and performance of the proposed method through various illustrative numerical examples.
	All experiments are implemented on a computer with Linux x86$\_$64 architecture and 100 cores.
	As common settings, the parameter settings used in the VAE training and the DDTD process are shown in Table~\ref{tbl:parameters}.

	\begin{table}[htbp]
		\caption{Parameter settings used in the VAE training and the DDTD process}
		\label{tbl:parameters}
		\begin{tabular*}{\tblwidth}{@{}LL@{}}
			\toprule
			\textbf{Parameter} & \textbf{Value} \\
			\midrule
			Mini-batch size (VAE)  & 20 \\
			Number of training epochs (VAE) & 100 \\
			Learning rate (VAE)  & $1.0 \times 10^{-4}$ \\
			Number of latent variables (VAE) & 8 \\
			KL divergence weight (DDTD) & 1 \\
			Maximum number of elite data (DDTD) & 400 \\
			\bottomrule
		\end{tabular*}
	\end{table}
	
	%%\begin{table}[]
	%%	\caption{Parameter settings used in the VAE training and the DDTD process}
	%%	\label{tbl:parameters}
	%%	\begin{tabular}{|c|c|}
		%%		\hline
		%%		\textbf{Parameter}                  & \textbf{Value}       \\ \hline
		%%		Mini-batch size (VAE)               & 20                   \\ \hline
		%%		Number of training epochs (VAE)     & 100                  \\ \hline
		%%		Learning rate (VAE)                 & $1.0 \times 10^{-4}$ \\ \hline
		%%		Number of latent variables (VAE)    & 8                    \\ \hline
		%%		KL divergence weight (DDTD)         & 1                    \\ \hline
		%%		Maximum number of elite data (DDTD) & 400                  \\ \hline
		%%	\end{tabular}
	%%\end{table}
	
	%%%\begin{figure}[ht]
	%%%	\begin {center}
	%%%	\includegraphics[width=1 \textwidth]{fig7.eps}
	%%%	\caption{Boundary conditions and design domains: \textbf{a} maximum von Mises stress (MVMS) minimization problem \textbf{b} compliant mechanism design problem \textbf{c} 3D optimization problem}
	%%%	\label{2D_boundaryCondition}
	%%%	\end {center}
	%%%\end{figure}
	
	\begin{figure}%[h]
		\begin {center}
		\includegraphics[width=1 \textwidth]{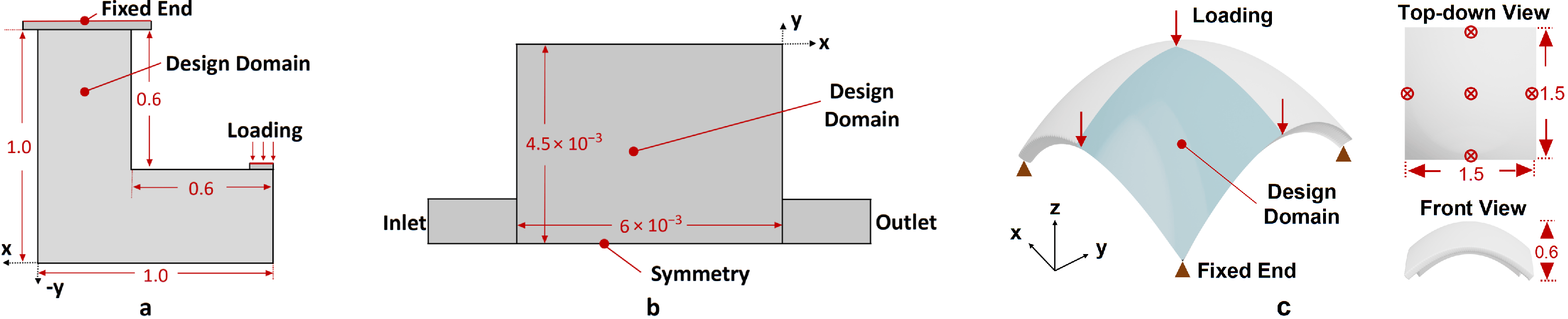}
		\caption{Boundary conditions and design domains: \textbf{a} maximum von Mises stress (MVMS) minimization problem \textbf{b} Microfluidic reactor design problem \textbf{c} Shell structural design problem.}
		\label{fig_design}
		\end {center}
	\end{figure}
	
	\subsection{Maximum von Mises stress minimization problem} \label{exam1}

	The stress-related L-bracket structure is first considered as a benchmark example to verify the effectiveness of the proposed method for strongly nonlinear optimization problems.
	Note that sensitivity-based TO methods often struggle to directly solve stress-related optimization problems.
	This difficulty mainly arises from the strong nonlinearity of stress responses and their highly localized nature, which often result in instability and poor convergence. 
	These difficulties are especially evident when minimizing the MVMS, as its sensitivity is hard to compute and often suffers from discontinuities and numerical instability.

	In this numerical example, the Young's modulus and Poisson's ratio are set to $1$ and $0.3$, respectively.
	The design domain and boundary conditions are shown in Figure~\ref{fig_design}(a).
	A total force of $2\times10^{-3}$ is uniformly distributed at the tip of the L-bracket to mitigate the stress concentration problem caused by concentrated loads.
	Geometrically nonlinear analysis is conducted to accurately capture the large-deformation response of the structure under the prescribed loading conditions.
	The design domain is discretized using quadrilateral finite elements with an edge length of 0.01.
	The filter radius $r^1$ in Eq.~\ref{eq2} is set to 0.03.
	The number of radii $N_{\mathrm{k}} $ in Eq.~\ref{eq_comp} is set to $10$.
	The maximum and minimum values of $R_k$ in Eq.~\ref{eq_comp} are set to $0.1$ and $0.03$, respectively.
	Moreover, the minimum weighting parameter $w_{\min}$ in Eq.~\ref{eq9} is chosen to be equal to the element edge length, i.e., $w_{\min}=0.01$.
	
	The maximum iteration number is $500$. 
	
	The following objective functions are considered to obtain L-bracket structures with low MVMS:
	\begin{equation}
		\label{eqE1}
		\begin{array}{ll}
			\underset{\boldsymbol{\rho} \in\{0,1\}}{\operatorname{Minimize}} & {\left[J_{1}( \phi(\boldsymbol{\rho}) ), J_{2}(  \phi(\boldsymbol{\rho}) )\right]} \\
			\text { where } & J_{1}( \phi(\boldsymbol{\rho}) ) = {\max}\left(\sigma^{v M}\right), \ J_{2}(\boldsymbol{\rho})= V.
		\end{array}
	\end{equation}
	where, $\phi(\boldsymbol{\rho})$ represents the material distribution reconstructed from $\boldsymbol{\rho}$.
	$\sigma^{v M}$ denotes the von Mises stress at an arbitrary point in the total analysis domain and $V$ denotes the volume of the total analysis domain.
	Note that, after extracting the iso-contour of $\phi = 0.5$, we make a body-fitted mesh in the solid domain ($\phi > 0.5$) and remove the void domain ($\phi < 0.5$).
	That is, the void domain is not included in the total analysis domain.
	
	\begin{figure}%[!t]
		\begin {center}
		\includegraphics[width=0.85 \textwidth]{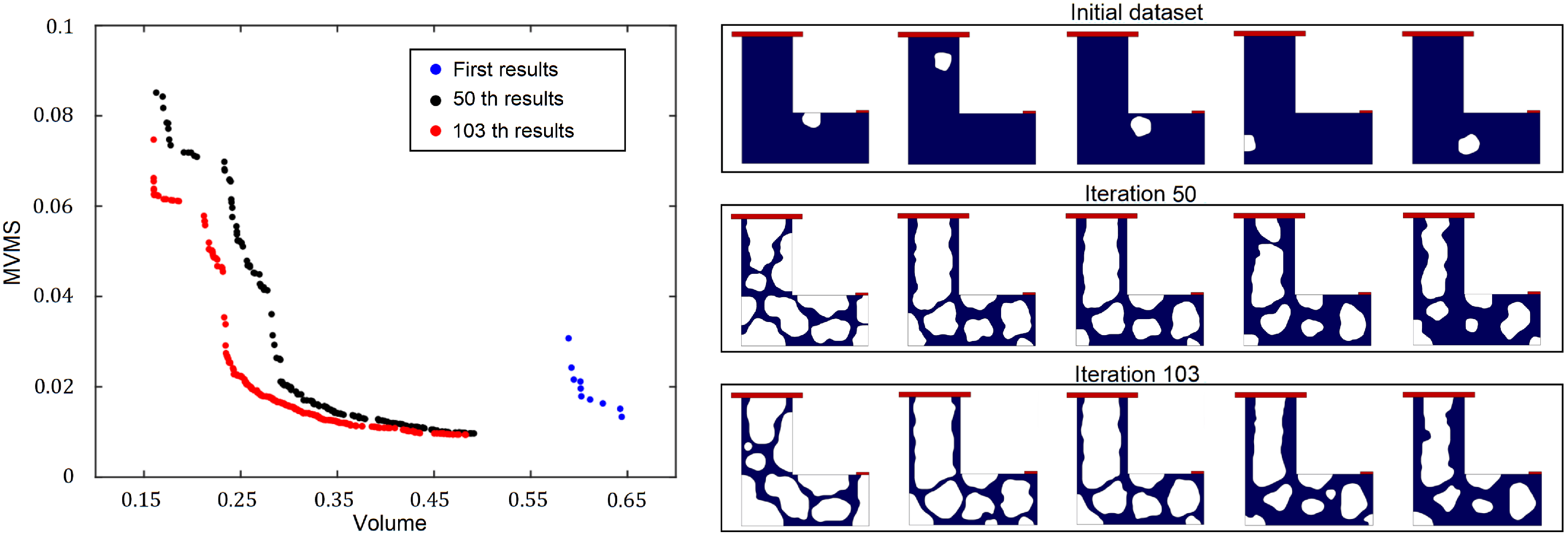}
		\caption{The improvement and partial results in the early stage of the L-bracket optimization problem.}
		\label{fig_ex1_early}
		\end {center}
	\end{figure}

	\begin{table}[]
		\caption{Average runtime and corresponding proportion of each part in a single iteration (unit: minute)}
		\label{tab:time_cost}
		\begin{tabular}{|c|cccccc|cccc|}
			\hline
			& \multicolumn{6}{c|}{Early Stage}                                                                                                                               & \multicolumn{4}{c|}{Later Stage}                                                                   \\ \hline
			Part       & \multicolumn{1}{c|}{VAE}    & \multicolumn{1}{c|}{Mutation} & \multicolumn{1}{c|}{RI}     & \multicolumn{1}{c|}{FEA}     & \multicolumn{1}{c|}{Others} & All   & \multicolumn{1}{c|}{VAE}     & \multicolumn{1}{c|}{FEA}     & \multicolumn{1}{c|}{Others}  & All   \\ \hline
			Time       & \multicolumn{1}{c|}{2.75}   & \multicolumn{1}{c|}{0.08}     & \multicolumn{1}{c|}{1.04}   & \multicolumn{1}{c|}{43.76}   & \multicolumn{1}{c|}{1.11}   & 48.74 & \multicolumn{1}{c|}{2.76}    & \multicolumn{1}{c|}{8.16}    & \multicolumn{1}{c|}{1.71}    & 12.63 \\ \hline
			Proportion & \multicolumn{1}{c|}{5.64\%} & \multicolumn{1}{c|}{0.17\%}   & \multicolumn{1}{c|}{2.13\%} & \multicolumn{1}{c|}{89.78\%} & \multicolumn{1}{c|}{2.28\%} & 100\% & \multicolumn{1}{c|}{21.85\%} & \multicolumn{1}{c|}{64.61\%} & \multicolumn{1}{c|}{13.54\%} & 100\% \\ \hline
		\end{tabular}
	\end{table}

\begin{figure}%[!t]
		\begin {center}
		\includegraphics[width=0.85 \textwidth]{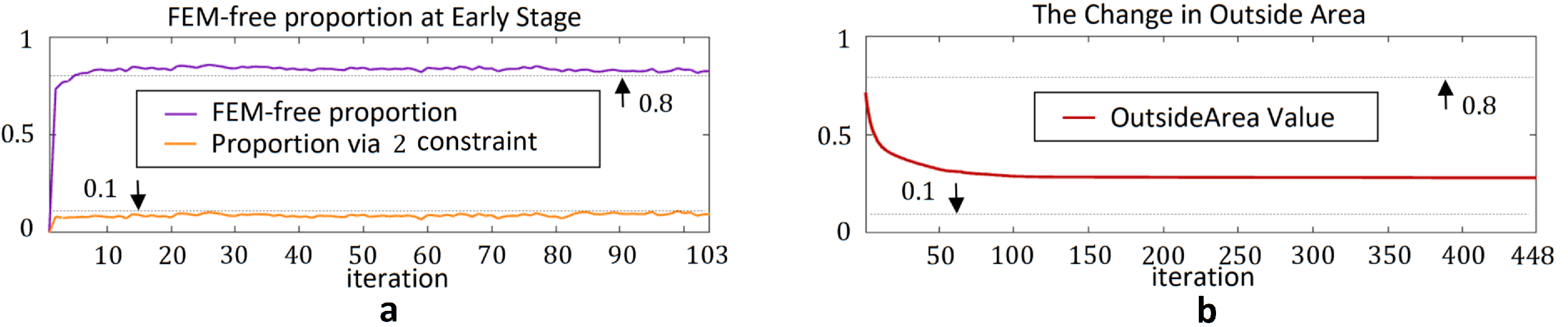}
		\caption{The proportion excluded from FEA calculations and the change in the outside area.}
		\label{fig_ex1_pro}
		\end {center}
	\end{figure}

	For this optimization problem, a low-information entropy initial dataset is constructed as introduced in Section~\ref{Data process flow}.
	After that, we selected $50$ initial material distributions as the initial dataset and part of them as shown in the first row on the right side of Figure~\ref{fig_ex1_early}.
	The transition condition between the early stage and later stage is defined as follows: if the change in the area outside of the elite data is below $5\times10^{-4}$ for $5$ consecutive iterations.
	The left side illustrates the improvement in the performance of elite data during the early stage of the DDTD iterative process.
	After being filtered by the RI algorithm and the 2 constraints in Section~\ref{minLength} only about 20\% of the newly generated structures proceed to FEM evaluation, while the remaining candidate data are discarded.
	It can be observed that the DDTD process can be driven by the low-information entropy initial dataset and generates material distributions with rough contours.
	Partial results from iterations~50 and 103 are shown on the right side.
	Notably, the results obtained at iteration 1 are not displayed on the right side.
	As discussed earlier, the rapid selection module enhances the computational efficiency of DDTD-based methods by significantly reducing the amount of data requiring FEA.
	Figure~\ref{fig_ex1_pro}(a) illustrates the proportion of data excluded from FEA evaluation to the total generated data (purple line) and the average proportion per iteration is 83.25\%.
	The proportion of data excluded from FEA evaluation pre-filtered through minimum length and connectivity constraints (yellow line) is also shown in the (a) and the average proportion per iteration is 8.90\%.
	The difference between these two proportions represents the ratio of potentially low-performance structures identified by the RI algorithm.
	The results indicate that the RI algorithm effectively eliminates around 74.35\% of the candidate structures from FEA evaluation (the number of structures in the realistic performance dataset
	$\hat{\boldsymbol{\rho}}_{\mathrm{real}}$ is set to 10\% of the total candidate data).
	It should be noted that the FEA evaluations of candidate structures in the DDTD process are completely independent. In practical applications, the computational efficiency of the proposed method can be further improved by performing these evaluations in parallel on multi-core CPUs.

	Furthermore, to further evaluate the impact of the proposed RI algorithm on the overall computational costs, the average runtime of each part during a single iteration of the DDTD process was recorded, as summarized in Table~\ref{tab:time_cost}. 
	The results show that FEA evaluation remains the dominant computational expense in the entire process, accounting for 89.78\% of the total runtime in early stage, with an average cost of 43.76 minutes per iteration. 
	In this case, a high-fidelity FEA evaluation of one structure requires approximately 1.81 minutes to ensure an accurate assessment of its structural performance.
	In contrast, the RI algorithm, which includes distance computation, MDS dimensionality reduction, and interpolation-based prediction, requires only 1.04 minutes on average, representing merely 2.13\% of the total runtime in early stage. 
	These results indicate that, although the RI algorithm introduces additional cost, its computational overhead is relatively small compared with the cost of FEA evaluations.
	Therefore, while maintaining the capability to identify potentially high-performance structures, the RI module effectively reduces the need for FEA evaluations for massive numbers of candidate structures, thereby significantly decreasing the overall computational cost. 
	Overall, the proposed RI algorithm replaces a massive amount of computationally expensive FEA calculations with a relatively small additional computational overhead, thus substantially improving the end-to-end computational efficiency of the DDTD process.

	\begin{figure}%[!t]
		\begin {center}
		\includegraphics[width=0.85 \textwidth]{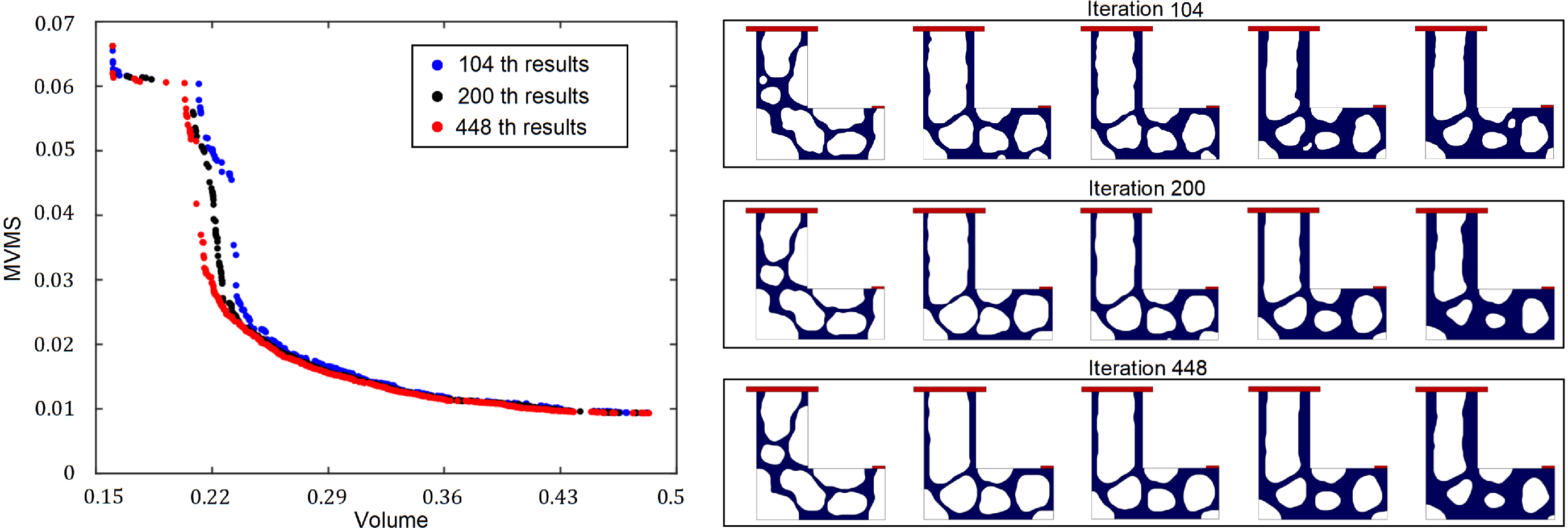}
		\caption{The improvement and partial results in the later stage of the L-bracket optimization problem.}
		\label{fig_ex1_Later}
		\end {center}
	\end{figure}
	
\begin{figure}%[!t]
		\begin {center}
		\includegraphics[width=0.85 \textwidth]{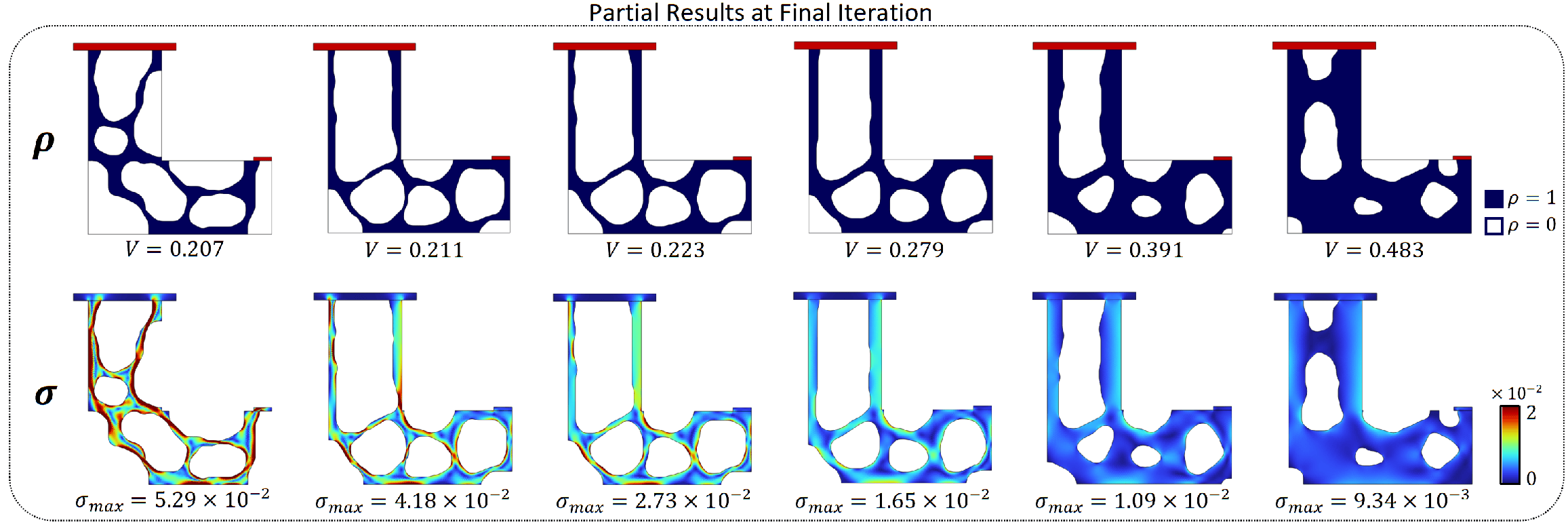}
		\caption{Partial final results and corresponding stress distributions.}
		\label{fig_ex1_results}
		\end {center}
	\end{figure}

	Figure~\ref{fig_ex1_Later} demonstrates the progressive improvement in the performance of elite data during the later stage of DDTD process.	
	In this VAE-dominated stage, the structural details are further refined, and the boundaries of the structures become increasingly continuous and smooth.
	Partial results for selected iterations are shown in the right side.
	The optimization process proceeds iteratively until the convergence criterion is satisfied, at which point the variation in the objective function becomes stable.
	In this case, the convergence criterion is defined as follows: if the change in the area outside of the elite data is below $1\times10^{-6}$ for $20$ consecutive iterations.
	The DDTD process reached convergence and was terminated after 448 iterations.
	The change in the area outside the elite data during the DDTD process is shown in Figure~\ref{fig_ex1_pro}(b).
	Figure~\ref{fig_ex1_results} displays a subset of elite material distributions and corresponding stress distributions at final iteration. 
	We obtained satisfactory elite data from the DDTD process driven by the low-information entropy initial dataset, demonstrating that the proposed method can effectively overcome the limitations faced by conventional DDTD-based methods, thereby significantly improving both effectiveness and generalizability.
	
	\begin{figure}%[!t]
		\begin {center}
		\includegraphics[width=0.85 \textwidth]{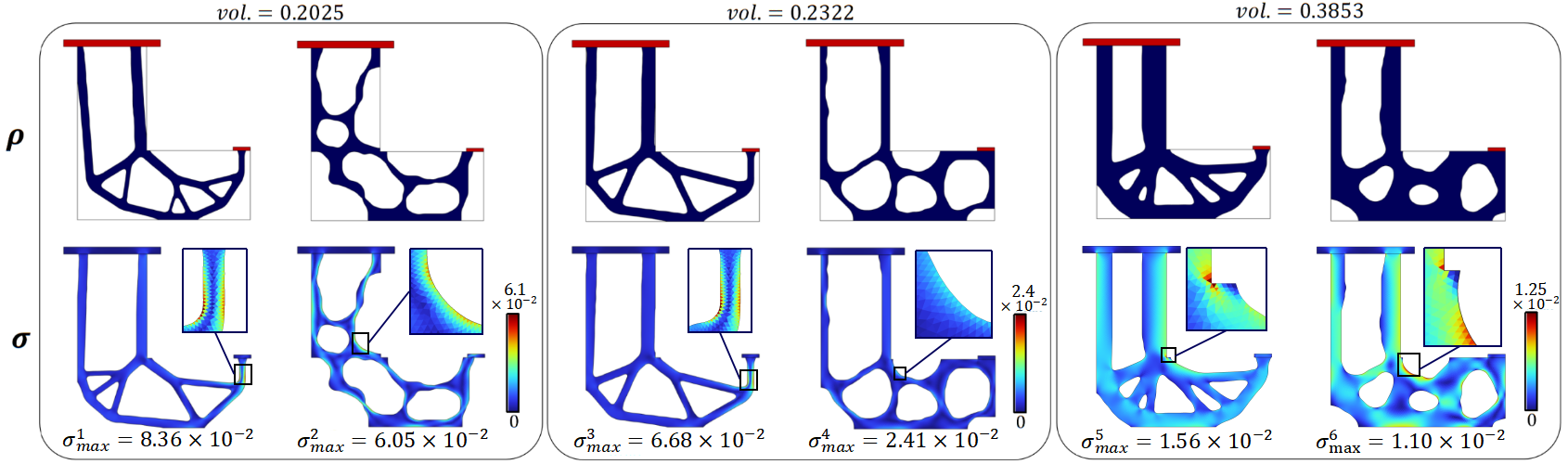}
		\caption{Comparison of the final results of DDTD with the results of sensitivity-based TO method.}
		\label{fig_ex1_comparsion}
		\end {center}
	\end{figure}

	Here, three structures from the elite data at the final iteration are selected for comparison with a sensitivity-based TO method implemented using the optimization module in the commercial multiphysics software COMSOL. 
	As illustrated in Figure~\ref{fig_ex1_comparsion}, each group of results consists of two columns. 
	The left column presents the material and stress distributions obtained by the sensitivity-based TO method, while the right column shows those obtained by the proposed method.
	Each set of comparison data was evaluated under similar volume (difference within $1\times 10^{-3}$) and the same body-fitted mesh generation condition.
	It can be observed that the DDTD results yield lower MVMS values (with DDTD results being $6.05$, $2.41$, $1.10\times 10^{-2}$, compared to the baseline values of $8.36$, $6.68$, $1.56\times 10^{-2}$), demonstrating that DDTD can achieve better results than the sensitivity-based TO method.
	These results demonstrate the effectiveness of the proposed method in improving structural performance for strongly nonlinear stress-related optimization problems.
	Notably, the sensitivity-based TO method used for comparison does not consider geometric nonlinearity, since incorporating large deformation effects into sensitivity-based TO method remains challenging. 
	As a result, the comparison is not completely fair when evaluating structures under large deformation conditions. 
	For the cases of vol = 0.2025 and vol = 0.2322, the structures obtained by the sensitivity-based TO method exhibit buckling in the magnified regions when subjected to geometrically nonlinear analysis, leading to significant stress concentrations. 
	In contrast, the structures generated by DDTD successfully avoid such buckling behavior and consequently achieve lower MVMS values. 
	For the case of vol = 0.3853, neither method exhibits buckling; however, DDTD still provides a superior design. 
	Although incorporating geometric nonlinearity into sensitivity-based TO remains challenging due to the difficulties associated with nonlinear equilibrium analysis, buckling behavior, and sensitivity computation, the proposed DDTD framework can directly handle such evaluations through high-fidelity simulations. These results demonstrate the effectiveness of DDTD for highly nonlinear structural optimization problems involving large deformation effects.

	\begin{figure}%[!t]
		\begin {center}
		\includegraphics[width=0.85 \textwidth]{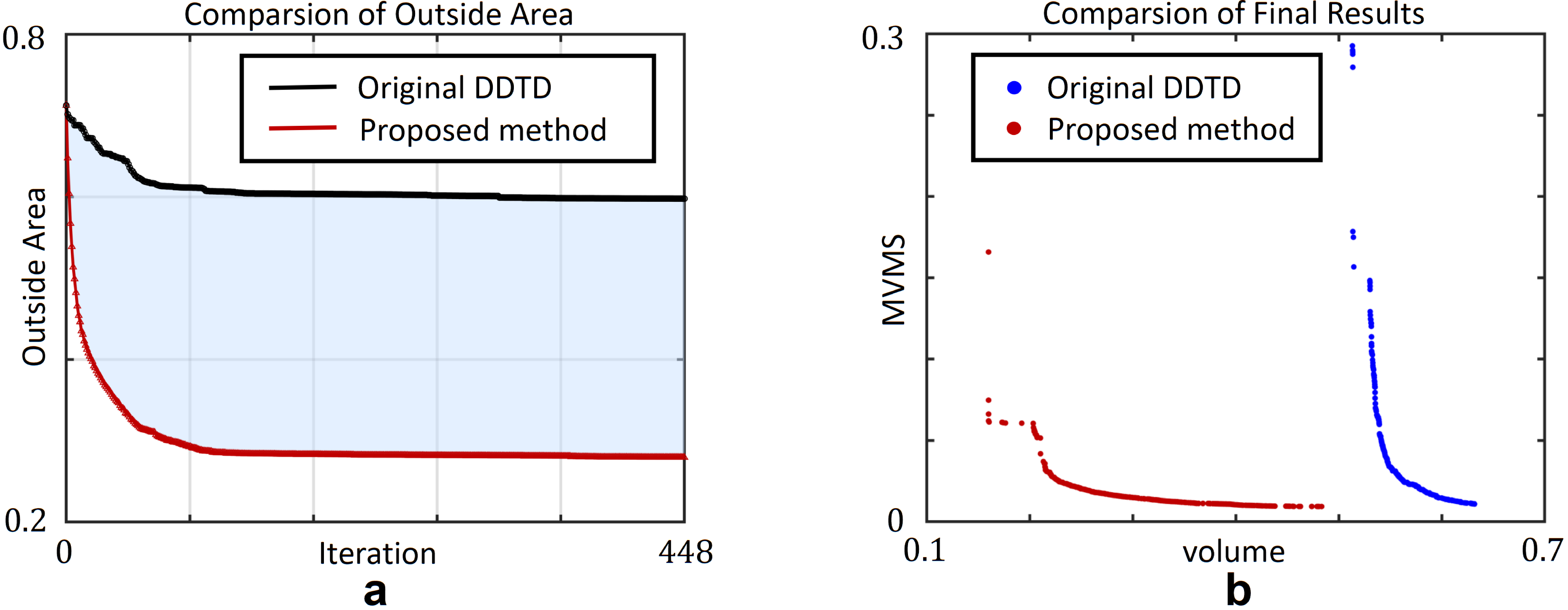}
		\caption{Comparison of conventional DDTD baseline with proposed method.}
		\label{fig_ex1_compareDDTD}
		\end {center}
	\end{figure}

	To validate the effectiveness of the proposed method compared with the conventional DDTD baseline (\cite{ref55}) under low-information entropy initial datasets, comparative experiments were conducted.
	For a fair comparison, both methods were implemented using the same initial dataset, simulation settings, and a similar total amount of data throughout the optimization process.

	As shown in Figure~\ref{fig_ex1_compareDDTD}(a), under the same low-information entropy initial dataset (with an initial Outside Area value of 0.7126), the proposed method ultimately converged to 0.2796, whereas the conventional DDTD baseline converged to 0.5980.
	Furthermore, the final results' performance of both methods are presented in Figure~\ref{fig_ex1_compareDDTD}(b).
	Compared with the conventional DDTD baseline, the proposed method achieved significantly superior structural performance, demonstrating stronger optimization capability and robustness under low-information entropy scenarios.

	DDTD employs unsupervised learning through deep generative models to generate diverse geometric structures while preserving key features of the training data.
	However, conventional DDTD-based methods heavily rely on deep generative models and therefore exist inherent limitation under low-information entropy conditions.

	\subsection{Microfluidic reactor design problem} \label{exam3}
	
	Compared with the previous structural optimization examples in Section~\ref{exam1}, this case further demonstrates the capability of the proposed method in microfluidic reactor design problem involving coupled transport and reaction processes.
	This problem focuses on optimizing the conversion efficiency of chemical reactions occurring in confined microscale flow environments.
	Microfluidic reactors typically consist of microscale channels and catalytic regions, where a carrier fluid transports reactants through the device and reactions occur on immobilized catalytic surfaces. 
	Due to the laminar flow conditions inherent at microscale conditions, mass transport limitations are significant, and system performance is strongly governed by the coupling between flow and catalyst layout.

	In this engineering problem, the goal is to determine the optimal catalyst layout that maximizes the mean conversion of reactants and minimizes the pressure drop.
	The catalyst layout influences not only the local reaction rate but also the global characteristics of the flow field, including velocity fields, pressure fields, and reactant concentration distributions. 
	Through an appropriate catalyst layout, the flow field can be guided to enhance mixing and establish an efficient balance between transport and reaction processes.
	This case represents a strongly nonlinear engineering problem involving the coupling among flow, diffusion, and chemical reaction.
	The boundary conditions and design domain are shown in Figure~\ref{fig_design}(b).
	The domains adjacent to the inlet and outlet boundaries are excluded from the design domain and are fixed as fluid throughout the optimization process.
	The design objective is to identify an optimal catalyst layout to maximize the mean reaction conversion and minimize the pressure drop $\Delta p$ as follows:
	
	\begin{equation}
		\label{eqEx3}
		\begin{array}{ll}
			\underset{\boldsymbol{\rho} \in\{0,1\}}{\operatorname{Minimize}} & {\left[J_{1}(\phi(\boldsymbol{\rho}) ), J_{2}( \phi(\boldsymbol{\rho}) )\right]} \\
			\text { where } & J_{1}( \phi(\boldsymbol{\rho}) ) = -\frac{1}{V_D} \int_{\Omega_c} r \ d \Omega, \ J_{2}( \phi(\boldsymbol{\rho}) )=\Delta p ,
			\\
			\text {subject to } & T_g \in \{N_{\mathrm{genus}}^{(1)}, N_{\mathrm{genus}}^{(2)}\}.
		\end{array}
	\end{equation}
	where, $V_D$ represents the volume of the design domain and $\Omega_c$ represents the catalyst domain ($\phi > 0.5$).
	$r$ is the reaction term of the first order isothermal reaction.
	In contrast to the numerical examples in Section~\ref{exam1}, we make a body-fitted mesh in both the catalyst domain $\Omega_c$ and the fluid domain $\Omega_f$ ($\phi < 0.5$).
	In this example, the design domain is discretized using quadrilateral finite elements with an element edge length of $5\times10^{-5}$.
	The filter radius $r^1$ in Eq.~\ref{eq2} is set to $1.5\times10^{-4}$.
	The number of radii $N_{\mathrm{k}} $ in Eq.~\ref{eq_comp} is set to $10$.
	The maximum and minimum values of $R_k$ in Eq.~\ref{eq_comp} are set to $1.5\times10^{-3}$ and $6\times10^{-4}$, respectively.
	Moreover, the minimum weighting parameter $w_{\min}$ in Eq.~\ref{eq9} is chosen to be equal to the element edge length, i.e., $w_{\min}=5\times10^{-5}$.
	
	The transition conditions of stage transition is the Outside Area value is below $2.5\times10^{-5}$ for $10$ consecutive iterations.
	The convergence criterion in this example is the change in the Outside Area value is below $1\times10^{-7}$ for $15$ consecutive iterations, or if the iteration count reaches the maximum iteration number $500$. 
	
	Notably, numerical simulations are performed for both the fluid domain and the catalyst domain in this case.
	Since the numerical analysis in the fluid domain is more prone to instability due to the solution of the Navier-Stokes equations, the minimum length and connectivity constraints described in Section~\ref{minLength} are applied to the fluid domain.

	$T_{g}$ is the genus value of fluid domain ($\phi < 0.5$) in one structure, and $N_{\mathrm{genus}}^{(1)}$, $N_{\mathrm{genus}}^{(2)}$ are the artificially defined constant values that are utilized to precisely control the topological complexities of the final solutions.
	The genus of one structure measures the maximum number of non-intersecting simple closed curves that can be drawn without separating it, effectively quantifying the number of holes or handles; for example, a sphere has a genus of 0, while a torus has a genus of 1~\citep{ref7}.
	The genus therefore serves as an indicator of the topological complexity of a structure. 
	Structures with excessive topological complexity are often difficult or even impossible to fabricate in practical manufacturing processes.
	In the 2D case, the genus value is equivalent to the number of enclosed cavities in one structure.
	
	In the implementation, the genus was calculated using the MATLAB built-in function 'bwboundaries', which extracts the boundaries of connected regions. 
	The detected boundaries consist of the structural outer boundary and the boundaries of enclosed cavities. 
	Therefore, the genus was evaluated by counting the number of enclosed cavities, which was obtained by subtracting the outer boundary from the total number of detected boundaries.

	\begin{figure}%[!t]
		\begin {center}
		\includegraphics[width=0.8 \textwidth]{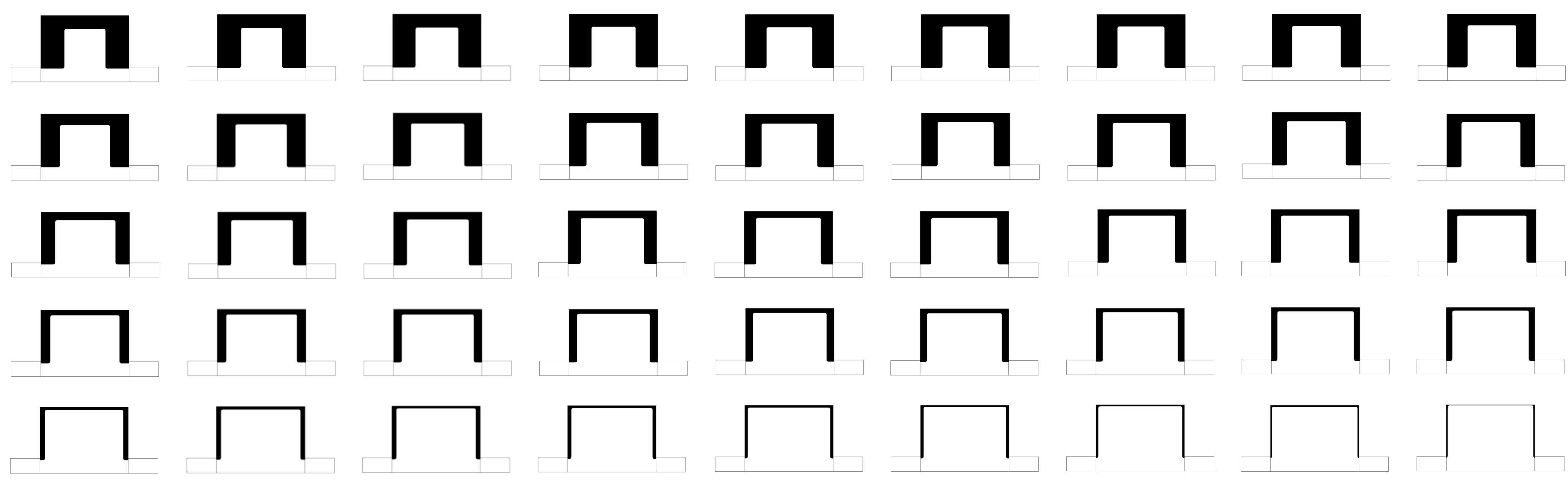}
		\caption{Low-information entropy initial dataset in microfluidic reactor design problem (the black area represents the catalyst domain, while the white area represents the fluid domain).}
		\label{fig_ex3_init}
		\end {center}
	\end{figure}

	As DDTD performs elite data selection according to the objective values of the structures, we transform the constrained optimization problem in Eq.~\ref{eqEx3} into an unconstrained formulation using the Lagrangian method (\cite{ref45}) as follows:
	\begin{equation}
		\label{ex3_genus_adj}
		\begin{array}{ll}
			\underset{\boldsymbol{\rho} \in\{0,1\}}{\operatorname{Minimize}} 
			& {\left[ \tilde{J}_{1}(J_{1},f_1), \tilde{J}_{2}(J_{2},f_1) \right]} \\
			\text{where } 
			& \tilde{J}_{1}=J_{1}+\lambda_{1} \, \ln(f_1 +1 ),\quad
			\tilde{J}_{2}=J_{2}+\lambda_{2} \, \ln(f_1 +1 ),\quad \\ &f_1=\min\left(\left|T_g-N_{\mathrm{genus}}^{(1)}\right|,\ \left|T_g-N_{\mathrm{genus}}^{(2)}\right|\right).
		\end{array}
	\end{equation}
	where, $\tilde{J_{1}}$ and $\tilde{J_{2}}$ are the equivalent optimization objectives of the original optimization problem in Eq.~\ref{eqEx3}, respectively.
	$\ln(*)$ denotes the natural logarithm. 
	The logarithmic transformation regularizes the penalty term $f_1$ and improves its continuity with respect to variations in $T_g$. 
	By applying the logarithm, abrupt changes in the magnitude of $f_1$ are attenuated, resulting in a smoother penalty landscape.
	$\lambda_{1}, \lambda_{2}$ are independent Lagrange multipliers, usually set to a constant value (in this numerical example, we use 100).
	The original optimization problem in Eq.~\ref{eqEx3} is solved equivalently by addressing the augmented objective formulation in Eq.~\ref{ex3_genus_adj}.

\begin{figure}%[!t]
		\begin {center}
		\includegraphics[width=0.9 \textwidth]{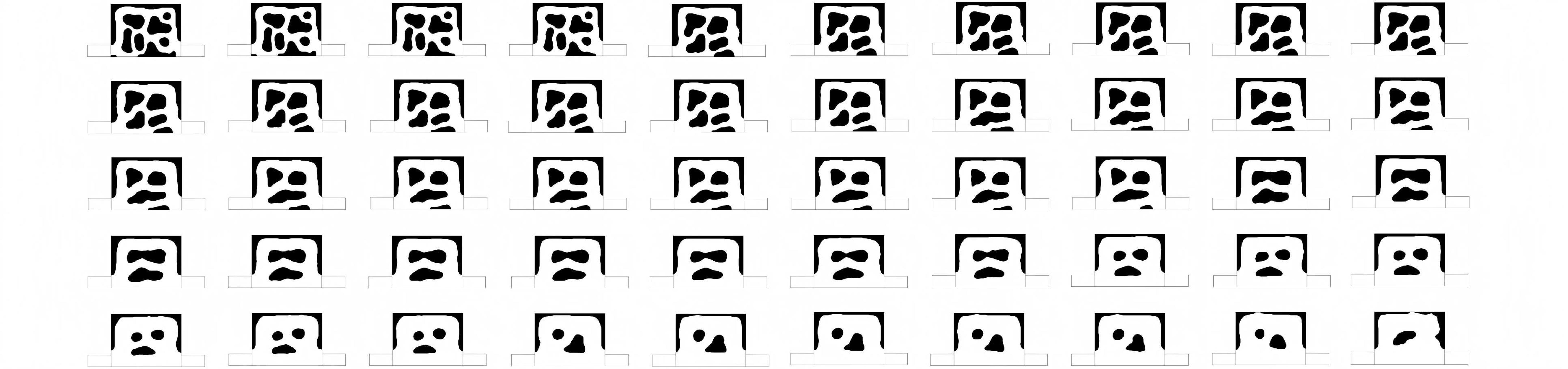}
		\caption{Partial results of the microfluidic reactor design problem without genus constraint.}
		\label{fig_ex3_noGenus}
		\end {center}
	\end{figure}

	To examine the robustness of the proposed method under different generation ways for low-information entropy initial datasets, a low-information entropy initial dataset is constructed, as shown in Figure~\ref{fig_ex3_init}, by prescribing the
	shapes of the fluid domain (white) and the catalyst domain (black).
	The microfluidic reactor design problem without the genus constraint is first investigated as a benchmark case before we present the effects of considering genus constraint on the geometry and topology of the structure.
	A subset of the final solutions under microfluidic reactor design problem without genus constraint is shown in Figure~\ref{fig_ex3_noGenus}. 
	The results clearly exhibit substantial variability in the value of $T_g$, with structures spanning from high topological complexity (e.g., $T_g=14$) to much simpler ones (e.g., $T_g=2$). 
	This wide range demonstrates that, in the absence of explicit topological constraint, the topological complexity of the optimized structures remains highly uncertain and can fluctuate significantly.

	\begin{figure}%[!t]
		\begin {center}
		\includegraphics[width=0.8 \textwidth]{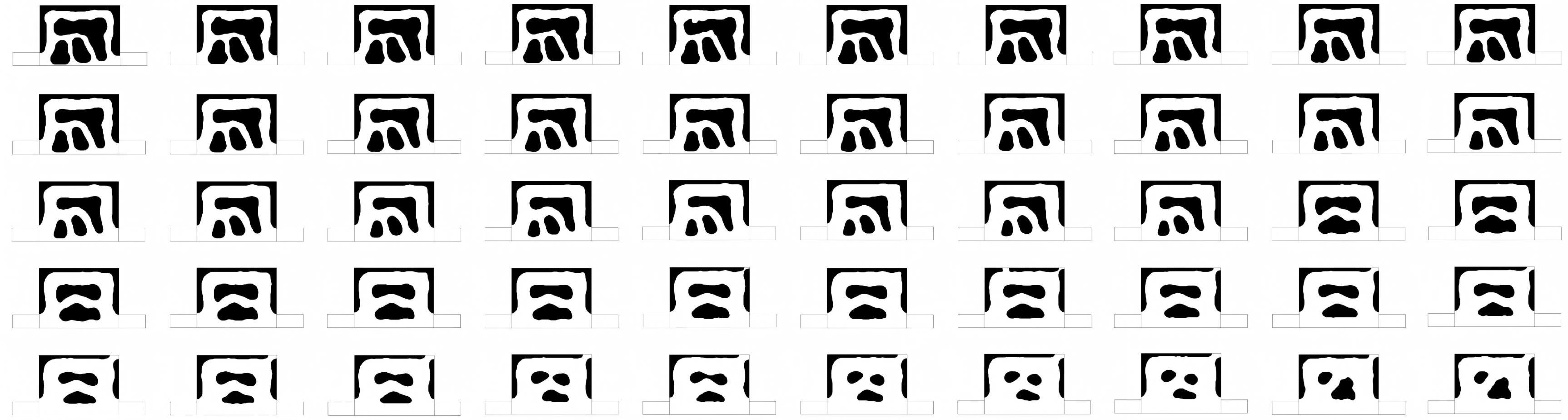}
		\caption{Partial results of the microfluidic reactor design problem with genus constraint. }
		\label{fig_ex3_genus}
		\end {center}
	\end{figure}

	In contrast, a subset of the final solutions obtained under the microfluidic reactor design problem with genus constraint (set $N_{\mathrm{genus}}^{(1)} = 4$ and $N_{\mathrm{genus}}^{(2)}=6$) is shown in Figure~\ref{fig_ex3_genus}. 
	All structures exhibit the same set topological complexity, and their geometries are significantly changed relative to those in Figure~\ref{fig_ex3_noGenus}. 
	This confirms that imposing a strict topological constraint effectively regulates the allowable structural complexity and substantially alters the material distribution.

	\begin{figure}%[!t]
		\begin {center}
		\includegraphics[width=0.8 \textwidth]{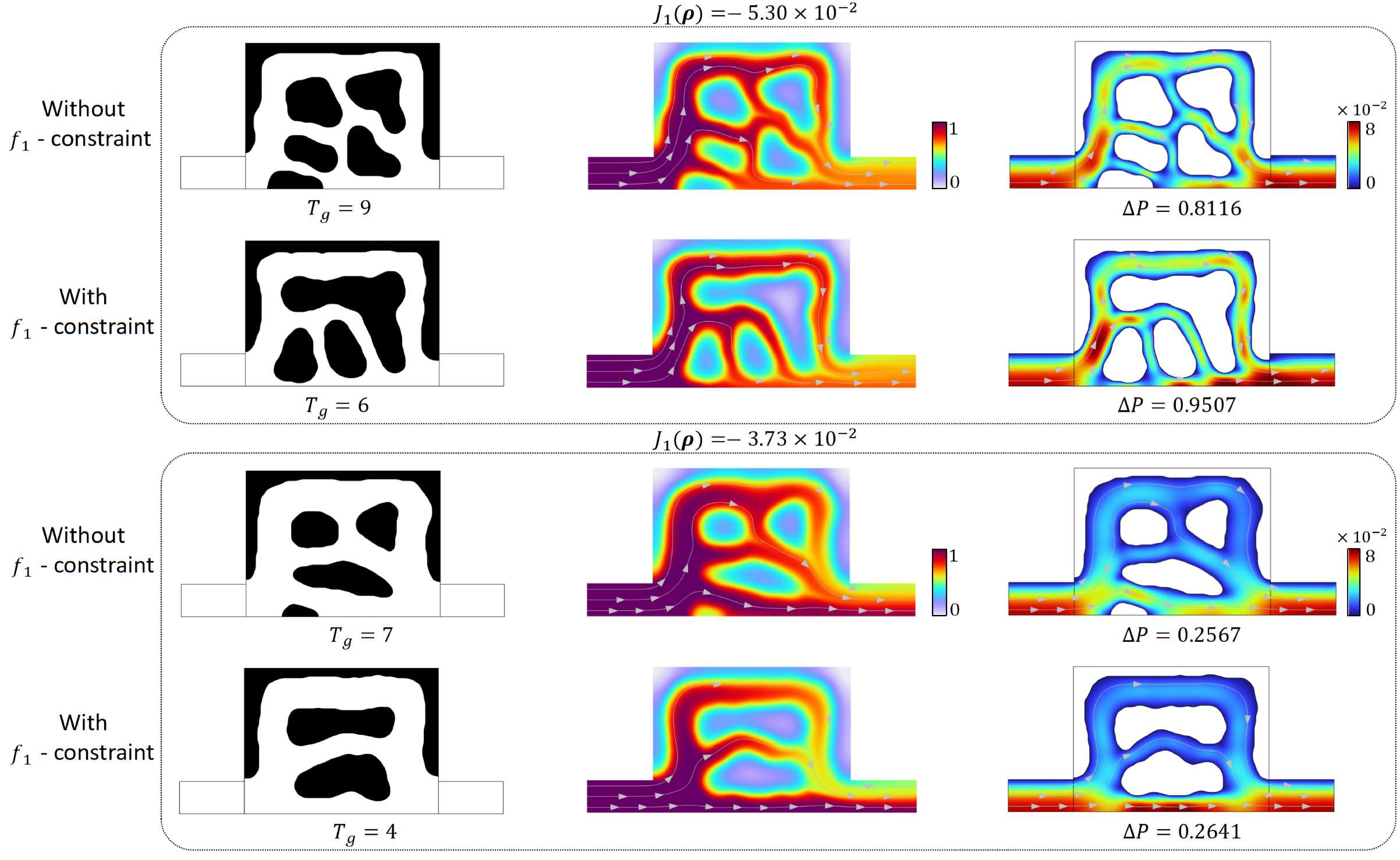}
		\caption{Comparison between the results of DDTD with/without genus constraint. }
		\label{fig_ex3_comparsion}
		\end {center}
	\end{figure}

	Figure~\ref{fig_ex3_comparsion} displays part of final elite data without genus constraint ($f_{1} -constraint$) and with genus restricted to 4 or 6. 
	To more intuitively compare the impact of the increasing genus constraints on the structural performance, we selected two groups of data under different genus constraints  (without and with constraint) for comparison, respectively. 
	The first group in Figure~\ref{fig_ex3_comparsion} shows the corresponding structures under different constraints at the same mean reaction conversion $5.30\times10^{-2}$ (differing less than $1\times10^{-4}$). 
	The corresponding pressure drops are $0.8116$, $0.9507$, and the corresponding topological complexities are $9$, $6$, respectively.
	Compared to the original problem without genus constraints, adding constraints limits the freedom of design, hence the decrease in the performance of the final structures is predictable.
	The second group shows the corresponding structures under different constraints at the same mean reaction conversion $3.73\times10^{-2}$ (differing less than $1\times10^{-4}$). 
	The corresponding pressure drops are $0.2567$, $0.2641$, and topological complexities are $7$, $4$, respectively.

	Sensitivity-based TO methods face inherent difficulties when a strict genus constraint is imposed. 
	The genus represents an integer-valued topological quantity, and changes in genus arise only through discrete topological events such as the creation or elimination of holes. 
	These events do not occur smoothly with respect to variations in the design variables, which prevents the genus from exhibiting a continuous or differentiable response. 
	Consequently, the strict equality constraint that enforces a prescribed genus is difficult to treat within sensitivity-based TO methods, and the problem becomes extremely difficult to solve using conventional sensitivity analysis.
	Additionally, due to the lack of sufficient prior information in this design problem, constructing a similar yet easy to solve pseudo-problem becomes challenging. 
	This issue makes conventional DDTD-based methods, which require a high-information entropy initial dataset, unable to construct a dataset that satisfies the prescribed topological condition, thereby rendering them inapplicable to this design problem.

	\subsection{Shell structural design problem} \label{exam2}

	Shell structures (\cite{ref_sup6}), characterized by their lightweight, structural stability, and aesthetic characteristics, are thin curved surfaces widely used in various engineering applications.
	An appropriate material distribution can effectively guide load transfer paths, enhance the global stiffness of the structure, and achieve lightweight design while maintaining sufficient load-carrying capacity.
	The boundary conditions and design domain are shown in Figure~\ref{fig_design}(c).
	In this numerical example, material properties were defined with Young's modulus of $1$ and Poisson's ratio of $0.3$. 
	Five vertical forces of $1\times 10^{-3}$ are applied at the input port.
	The design domain is discretized using quadrilateral finite elements with an element edge length of 0.005.
	The filter radius $r^1$ in Eq.~\ref{eq2} is set to 0.045.
	The number of radii $N_{\mathrm{k}} $ in Eq.~\ref{eq_comp} is set to $10$.
	The maximum and minimum values of $R_k$ in Eq.~\ref{eq_comp} are set to $0.1$ and $0.03$, respectively.
	Moreover, the minimum weighting parameter $w_{\min}$ in Eq.~\ref{eq9} is set to $0.01$ in order to avoid generating extremely thin hinge-like regions, in addition to suppressing excessively detailed geometric features.
	
	The transition conditions of stage transition is the Outside Area value is below $5\times10^{-4}$ for $5$ consecutive iterations.
	The convergence criterion in this example is the change in the Outside Area value is below $1\times10^{-6}$ for $20$ consecutive iterations, or if the iteration count reaches the maximum iteration number $150$. 
	
	The optimization problem considering genus constraint is defined as follows:
	\begin{equation}
		\label{eqExShell}
		\begin{array}{ll}
			\underset{\phi(\boldsymbol{\rho}) \in\{0,1\}}{\operatorname{Minimize}} 
			& {\left[J_{1}(\phi(\boldsymbol{\rho})), J_{2}(\phi(\boldsymbol{\rho}))\right]} \\ [5pt]
			\text{where } 
			& J_{1}(\phi(\boldsymbol{\rho})) 
			= \displaystyle \int_{\Omega} \boldsymbol{f} \cdot \boldsymbol{u} \, \mathrm{d}\Omega,\ J_{2}(\phi(\boldsymbol{\rho})) = V, \\ [5pt]
			\text{subject to } 
			& T_g =N_{\mathrm{genus}}.
		\end{array}
	\end{equation}
	where, $\boldsymbol{f}$ denotes the external load vector applied to the shell structure, representing the prescribed mechanical actions acting on $\Omega$, and $\mathbf{u}$ denotes the displacement vector.
	$V$ denotes the volume of the body-fitted material distribution $\phi(\boldsymbol{\rho})$.
	$N_{\mathrm{genus}}=4$ is an artificially defined constant value.
	The optimization objectives of Eq.~\ref{eqExShell} are to minimize the overall structural compliance ($J_{1}$), thereby improving load-bearing performance while simultaneously promoting structural lightweighting.
	
	\begin{figure}%[!t]
		\begin {center}
		\includegraphics[width=0.9 \textwidth]{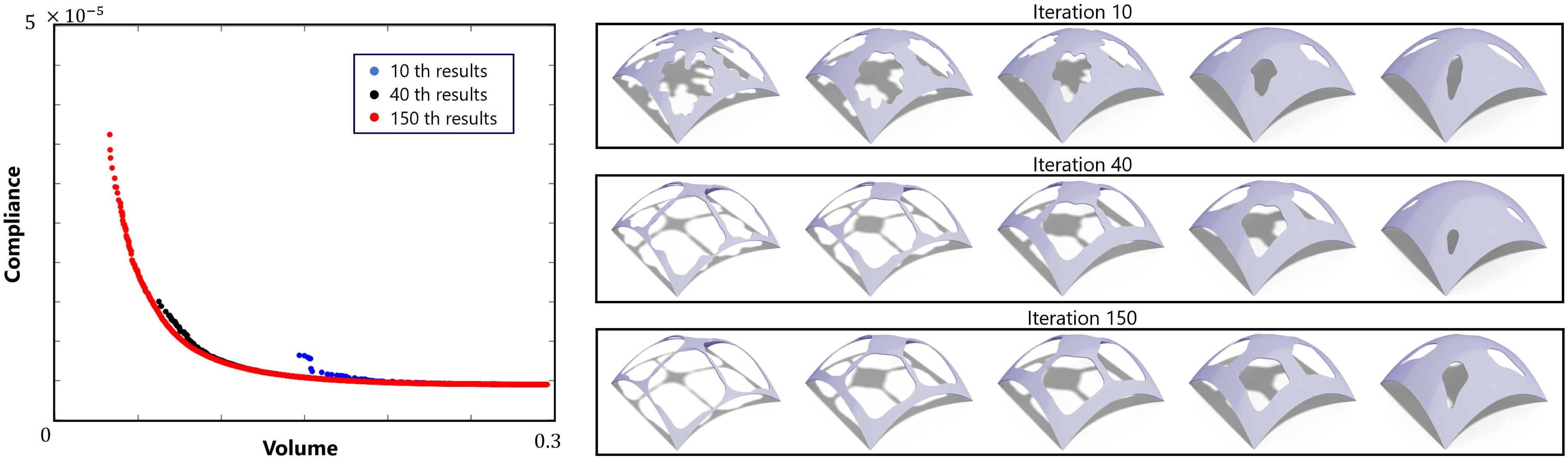}
		\caption{The improvement and partial results of the shell structural design problem with genus constraint.}
		\label{fig_ex2}
		\end {center}
	\end{figure}

	\begin{figure}%[!t]
		\begin {center}
		\includegraphics[width=0.9 \textwidth]{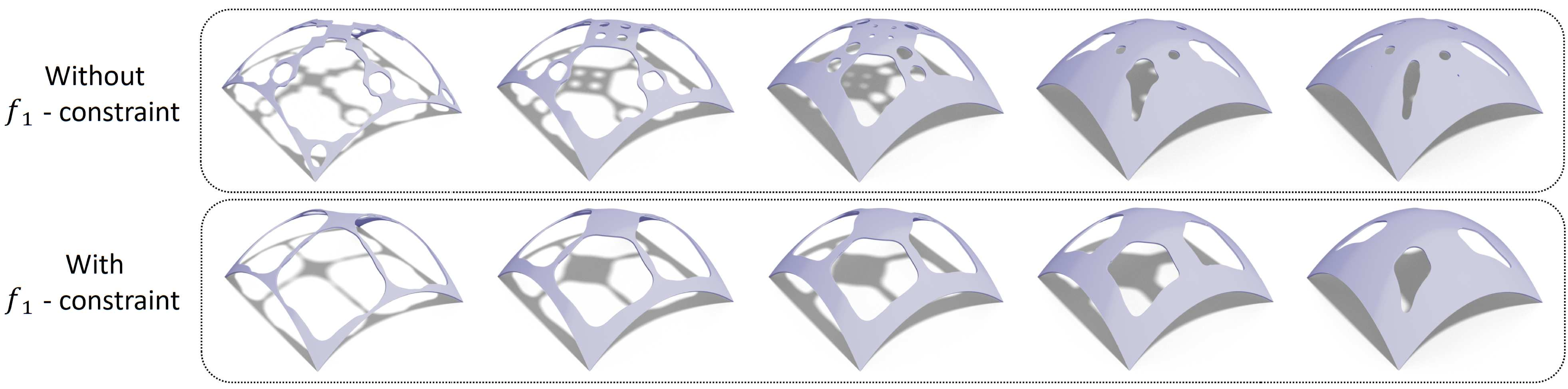}
		\caption{Comparison of results obtained from the shell structural design problems with and without genus constraints.}
		\label{fig_ex2_comp}
		\end {center}
	\end{figure}
	
	Similar to Section~\ref{exam3}, we transform the constrained optimization problem in Eq.~\ref{eqExShell} into an unconstrained formulation.
	We constructed a low-information entropy initial dataset as introduced in Section~\ref{Data process flow}.
	The DDTD process reached convergence and was terminated after $150$ iterations.
	The left side of Figure~\ref{fig_ex2} illustrates the improvement in performance of elite data during the DDTD iterative process. 
	Due to the inclusion of two optimization objectives, structural compliance and volume, within this optimization problem, the enhancement of elite data progresses toward balancing the optimization objectives.
	Partial results from selected iterations are shown on the right side.
	As seen, satisfactory elite data can be obtained from the DDTD process driven by the low-information entropy initial dataset. 
	In Figure~\ref{fig_ex2_comp}, the shape differences between the results of the shell structural design problem without genus constraints and those with genus constraints are compared. 
	It can be observed that the proposed method effectively controls the genus.
	This example further demonstrates that the proposed framework can effectively address optimization problems involve non-differentiable characteristics, which are difficult to handle using sensitivity-based TO methods as well as conventional DDTD-based methods.
	
	%% Displayed equations can be tagged using various environments. 
	%% Single line equations can be tagged using the equation environment.
	
	%\begin{table}[t]%% placement specifier
	%% Use tabular environment to tag the tabular data.
	%% https://en.wikibooks.org/wiki/LaTeX/Tables#The_tabular_environment
	%\centering%% For centre alignment of tabular.
	%\begin{tabular}{l c r}%% Table column specifiers
	%% Tabular cells are separated by &
	% 1 & 2 & 3 \\ %% A tabular row ends with \\
	%  4 & 5 & 6 \\
	%  7 & 8 & 9 \\
	%\end{tabular}
	%% Use \caption command for table caption and label.
	%\caption{Table Caption}\label{fig1}
	%\end{table}
	
	\section{Conclusion}
	\label{sec4}
	
	This study presented a rapid identification based DDTD-based framework that remains valid under low-information entropy scenarios. 
	A mesh-independent, parameter-controllable mutation module was proposed to generate geometric features avoiding over-reliance on deep generative model, and a SDF-based minimum length constraint together with a connectivity constraint ensured stable generation of a body-fitted mesh.
	The embedding-based rapid identification algorithm identifies potential high-performance structures directly in a physics-informed embedding space and substantially reduces the ratio of data requiring numerical simulation. 
	
	The stress-related experiments verified the effectiveness and robustness of the proposed framework. 
	The microreactor design problem with strict genus constraint further demonstrated that the method can address non-differentiable optimization problems that are difficult for both sensitivity-based TO and conventional DDTD-based methods.
	These results confirm the capability of the proposed framework to handle strongly nonlinear physics, strict topological constraints, and limited prior information.
	
	Future research will focus on further improving the practical applicability and computational efficiency of the proposed framework. 
	In particular, additional manufacturability constraints for 3D printing, such as self-supporting and connectivity constraints, will be incorporated into the optimization process to improve the manufacturability of the generated structures. Furthermore, GPU-based acceleration strategies and alternative high-efficiency numerical simulation approaches, such as physics-informed neural networks (PINNs), will be investigated to further accelerate the high-fidelity numerical simulation process and improve the scalability of the framework for large-scale TO problems.

	% Uncomment and use as the case may be
	%\begin{theorem} 
	%\end{theorem}
	
	% Uncomment and use as the case may be
	%\begin{lemma} 
	%\end{lemma}
	
	
	\section{}\label{}
	
	\printcredits
	
	\bibliographystyle{cas-model2-names}
	%\bibliographystyle{cas-model2-names}
	%\bibliographystyle{elsarticle-num}

	\bibliography{cas-refs}

	
	
\end{document}